\documentclass[12pt]{article}
\usepackage{amsmath,amssymb,amsthm,amsxtra,overpic,bbm,bm,epsfig}
\usepackage{hyperref}
\usepackage{mathrsfs}
\usepackage{enumitem}
\usepackage{graphicx}
\usepackage{color}
\usepackage[table]{xcolor}
\usepackage{comment}
\usepackage{epstopdf}
\usepackage{float}
\usepackage{cite}
\usepackage{slashed,stmaryrd}
\usepackage{youngtab}
\usepackage{makecell}

\usepackage{diagbox}
\usepackage{blkarray}
\usepackage{tabularx}
\usepackage{pinlabel}

\usepackage[title]{appendix}

\allowdisplaybreaks[2]

\def\thefootnote{\fnsymbol{footnote}}

\hypersetup{
	colorlinks =true,
	linkcolor  = red,
	citecolor  = blue,
	urlcolor   = blue
}

\usepackage{multirow}
\usepackage{mathtools}
\usepackage{subcaption}

\usepackage{ulem}

\newcommand{\rmi}{{\rm i}}

\begin{document}
	
%	{\normalsize \flushright TUM-HEP 1460/23\\}
%	\vspace{0.2cm}
	
%\title{\vspace{-2cm}
%{\normalsize \flushright TUM-HEP 1447/22, IPMU22-0070\\}
%\vspace{-1.5cm}}
%
%\date{}
%\maketitle

\begin{center}
{\Large\bf Trimaximal Mixing Patterns Meet the First  JUNO Result}
\end{center}

%\vspace{0.2cm}

\begin{center}
{\bf  Di Zhang}~\footnote{E-mail: di1.zhang@tum.de} %\;, \quad {\bf~$^{a,~c}$}~\footnote{}
\\
\vspace{0.2cm}
{\small
Physik-Department, Technische Universität München, James-Franck-Straße, 85748 Garching, Germany}
\end{center}

%\vspace{1.5cm}

\begin{abstract}
	
The JUNO experiment has recently released its first measurement results based on 59.1 days of data, achieving unprecedented precision in measuring the lepton mixing angle $\theta_{12}$. This significant improvement places stringent constraints on certain neutrino mass models and flavor mixing patterns. In this work, we examine the impact of the latest JUNO results on the two trimaximal (i.e., TM1 and TM2) mixing patterns. They are two well-motivated variants of the tri-bimaximal mixing pattern and predict specific correlations between $\theta_{12}$ and $\theta_{13}$.  After taking into account the first JUNO results, the TM1 mixing pattern sits on the edge of the experimentally allowed $1\sigma$ region, while the TM2 mixing pattern lies outside the $3\sigma$ region. To reconcile these TM mixing patterns with the latest experimental data, we further investigate the renormalization group (RG) running effects on them in the both Majorana and Dirac neutrino cases. Our analytical and numerical results show that RG corrections can bring the two TM mixing patterns into excellent agreement with the latest JUNO data if neutrino masses are quasi-degenerate. However, the Majorana case faces severe constraints from neutrinoless double beta decay limits, and particularly, the TM2 mixing pattern with Majorana neutrinos has been essentially ruled out. In the Dirac case, the TM1 mixing pattern is fully consistent with current data including beta decay results, whereas the TM2 pattern is strongly constrained by the KATRIN limit and even could be largely ruled out if the KATRIN experiment reaches its final sensitivity without any discovery. Future high-precision measurements of lepton mixing parameters and absolute neutrino masses in both oscillation and non-oscillation experiments will provide decisive tests of these mixing patterns.

\end{abstract}

%\begin{flushleft}
%\hspace{0.8cm} PACS number(s):
%\end{flushleft}

\def\thefootnote{\arabic{footnote}}
\setcounter{footnote}{0}

\newpage
%\tableofcontents

\section{Introduction} \label{sec:Introduction}

Neutrino oscillation experiments have firmly established that lepton flavors are mixed through two large and one small mixing angles, in stark contrast to the small mixing observed in the quark sector~\cite{ParticleDataGroup:2024cfk,Xing:2020ijf}. This peculiar pattern of lepton flavor mixing has inspired numerous neutrino mass models based on flavor symmetries, many of which predict simple and appealing constant mixing patterns (see, e.g., reviews~\cite{Altarelli:2010gt,Ishimori:2010au,King:2013eh,Xing:2015fdg,Petcov:2017ggy,Feruglio:2019ybq,Xing:2022uax,Chauhan:2023faf,Ding:2023htn,Ding:2024ozt} and references therein). One of them is the tri-bimaximal (TBM) mixing pattern~\cite{Harrison:2002er,Xing:2002sw,He:2003rm}:
\begin{eqnarray}
	U^{}_{\rm TBM} = \begin{pmatrix} \dfrac{2}{\sqrt{6}} & \dfrac{1}{\sqrt{3}} & 0 \\[0.35cm] -\dfrac{1}{\sqrt{6}} & \dfrac{1}{\sqrt{3}} & \dfrac{1}{\sqrt{2}}  \\[0.35cm] \dfrac{1}{\sqrt{6}} & - \dfrac{1}{\sqrt{3}} & \dfrac{1}{\sqrt{2}}  \end{pmatrix} \;,
\end{eqnarray}
predicting $\theta^{}_{23} = 45^\circ$, $\theta^{}_{12} = \arctan\left( \sqrt{2}/2 \right) = 35.26^\circ$, and $\theta^{}_{13} = 0$ under the standard parametrization~\cite{ParticleDataGroup:2024cfk} of the Pontecorvo-Maki-Nakagawa-Sakata (PMNS) matrix~\cite{Pontecorvo:1957cp,Maki:1962mu,Pontecorvo:1967fh},
\begin{eqnarray}\label{eq:parametrization}
	U = \begin{pmatrix} c^{}_{12} c^{}_{13} & s^{}_{12} c^{}_{13}  & s^{}_{13} e^{-\rmi \delta} \\ -s^{}_{12} c^{}_{23} - c^{}_{12}s^{}_{13} s^{}_{23} e^{\rmi\delta} & c^{}_{12} c^{}_{23} - s^{}_{12} s^{}_{13}s^{}_{23} e^{\rmi \delta} & c^{}_{13} s^{}_{23} \\ s^{}_{12} s^{}_{23} - c^{}_{12} s^{}_{13} c^{}_{23} e^{\rmi\delta} & -c^{}_{12} s^{}_{23} - s^{}_{12} s^{}_{13} c^{}_{23} e^{\rmi\delta} & c^{}_{13} c^{}_{23} \end{pmatrix} P^{}_\nu \;,
\end{eqnarray}
where $c^{}_{ij} = \cos\theta^{}_{ij}$ and $s^{}_{ij} = \sin\theta^{}_{ij}$ (for $ij=12,13,23$), and $P^{}_\nu = {\rm Diag} \{ e^{\rmi \rho}, e^{\rmi\sigma}, 1 \}$ is the Majorana (or unphysical) phase matrix in the Majorana (or Dirac) neutrino case. The former two predictions are very close to the best-fit values of $\theta^{}_{23}$ and $\theta^{}_{12}$ and lie within their $3\sigma$ ranges~\cite{Esteban:2024eli,Capozzi:2025wyn}. This success had motivated extensive studies on flavor-symmetry realizations and phenomenological consequences of the TBM mixing pattern. However, $\theta^{}_{13}$ turns out to be not as tiny as initially expected~\cite{T2K:2011ypd,DoubleChooz:2011ymz,DayaBay:2012fng,RENO:2012mkc}, and the most precise measurement from the Daya bay experiment gives $ \sin^22\theta^{}_{13} = 0.0851 \pm 0.0024$~\cite{DayaBay:2022orm}, which strongly disfavors the TBM mixing pattern and calls for its modifications or alternatives. 

The trimaximal (TM) mixing patterns are the simplest and most popular variants of the TBM pattern. They are obtained by introducing a single complex rotation to the right of $U^{}_{\rm TBM}$, such that one column of $U^{}_{\rm TBM}$ remains untouched. A rotation in the 1-2 plane leaves $\theta^{}_{13}$ vanishing, making it phenomenologically uninteresting. In contrast, the rotations in the 2-3 and 1-3 planes introduced to the right of $U^{}_{\rm TBM}$ are able to generate a non-vanishing $\theta^{}_{13}$ depending on the rotation angle $\theta$, and are thus much more attractive. These two variants are usually referred to as the TM1 and TM2 mixing patterns~\cite{Harrison:2002kp,Xing:2006xa,Xing:2006ms,Lam:2006wm,Grimus:2008tt,Albright:2008rp}, respectively, and are explicitly given by
\begin{eqnarray}\label{eq:TM1}
	U^{}_{\rm TM1} = \begin{pmatrix} \dfrac{2}{\sqrt{6}} & \dfrac{\cos\theta}{\sqrt{3}} &\dfrac{\sin\theta}{\sqrt{3}} e^{-\rmi \phi} \\[0.35cm] - \dfrac{1}{\sqrt{6}} & \dfrac{\cos\theta}{\sqrt{3}} - \dfrac{\sin\theta}{\sqrt{2}} e^{\rmi \phi} & \dfrac{\cos\theta}{\sqrt{2}} + \dfrac{\sin\theta}{\sqrt{3}} e^{-\rmi \phi}  \\[0.35cm] \dfrac{1}{\sqrt{6}} & - \dfrac{\cos\theta}{\sqrt{3}} - \dfrac{\sin\theta}{\sqrt{2}} e^{\rmi \phi} & \dfrac{\cos\theta}{\sqrt{2}} - \dfrac{\sin\theta}{\sqrt{3}} e^{-\rmi \phi}  \end{pmatrix} \;,
\end{eqnarray}
and
\begin{eqnarray}\label{eq:TM2}
	U^{}_{\rm TM2} = \begin{pmatrix} \dfrac{2\cos\theta}{\sqrt{6}} & \dfrac{1}{\sqrt{3}} & \dfrac{2\sin\theta}{\sqrt{6}} e^{-\rmi \phi} \\[0.35cm] -\dfrac{\cos\theta}{\sqrt{6}} -  \dfrac{\sin\theta}{\sqrt{2}} e^{\rmi \phi} & \dfrac{1}{\sqrt{3}} & \dfrac{\cos\theta}{\sqrt{2}} - \dfrac{\sin\theta}{\sqrt{6}} e^{-\rmi \phi} \\[0.35cm] \dfrac{\cos\theta}{\sqrt{6}} - \dfrac{\sin\theta}{\sqrt{2}} e^{\rmi \phi}  & - \dfrac{1}{\sqrt{3}} & \dfrac{\cos\theta}{\sqrt{2}} + \dfrac{\sin\theta}{\sqrt{6}} e^{-\rmi \phi} \end{pmatrix} \;.
\end{eqnarray}
The mixing angle $\theta^{}_{13}$ in these two trimaximal patterns are determined by the rotation angle $\theta$, namely, $\sin\theta^{\rm TM1}_{13} =\sin\theta/\sqrt{3}$ and $\sin\theta^{\rm TM2}_{13} =2\sin\theta/\sqrt{6}$. Furthermore, an intriguing correlation between $\theta^{}_{12}$ and $\theta^{}_{13}$ in each pattern can be easily achieved from Eqs.~\eqref{eq:TM1} and \eqref{eq:TM2}~\footnote{In a recent paper~\cite{Xing:2025bdm}, the correlation between $\theta^{}_{12}$ and $\theta^{}_{13}$ in the TM1 mixing pattern has been generalized into that among the three elements in the first row of $U$, namely $|U^{}_{e1}|^2 = 2\left( |U^{}_{e2}|^2 + | U^{}_{e3}|^2 \right) $. This relation at the matrix element level is remarkably applicable to a non-unitary $U$. The potential experimental tests of the relation and non-unitarity with the JUNO and Daya~Bay measurements have also been discussed.}:
\begin{eqnarray}\label{eq:TM-relations}
	\sin^2\theta^{\rm TM1}_{12} = \frac{1}{3} - \frac{2}{3}\tan^2\theta^{\rm TM1}_{13} \;,\quad \sin^2\theta^{\rm TM2}_{12} = \frac{1}{3} +\frac{1}{3}\tan^2\theta^{\rm TM2}_{13} \;.
\end{eqnarray}

\begin{table}[t!]
	\centering
	\renewcommand{\arraystretch}{1.3}
	\resizebox{\textwidth}{!}{
	\begin{tabular}{c|c|c|c|c|c|c|c}
		\hline\hline
		\rule{0pt}{0.72cm}& & $\sin^2\theta^{}_{12}$ & $\sin^2\theta^{}_{13}$ & $\sin^2\theta^{}_{23}$ & $\delta/^\circ$ & $\dfrac{\Delta m^2_{21}}{10^{-5}~{\rm eV}^2}$ & $\dfrac{\Delta m^2_{3\ell}}{10^{-3}~{\rm eV}^2}$
		\\[0.2cm]
		\hline
		\multirow{2}{*}{NuFIT 6.0} & NMO & \multirow{2}{*}{$0.308^{+0.012}_{-0.011}$} & $0.02215^{+0.00056}_{-0.00058}$ & $0.470^{+0.017}_{-0.013}$ & $212^{+26}_{-41}$ & \multirow{2}{*}{$7.49^{+0.19}_{-0.19}$} & $+2.513^{+0.021}_{-0.019}$ 
		\\
		& IMO &  & $0.02231^{+0.00056}_{-0.00056}$ & $0.550^{+0.012}_{-0.015}$ & $274^{+22}_{-25}$ & & $-2.484^{+0.020}_{-0.020}$ 
		\\
		\hline
		\multirow{2}{*}{NuFIT 6.1} & NMO & \multirow{2}{*}{$0.3088^{+0.0067}_{-0.0066}$} & $0.02248^{+0.00055}_{-0.00059}$ & $0.470^{+0.017}_{-0.014}$ & $212^{+26}_{-36}$ & \multirow{2}{*}{$7.537^{+0.094}_{-0.10}$} & $+2.511^{+0.021}_{-0.020}$ 
		\\
		& IMO &  & $0.02262^{+0.00057}_{-0.00056}$ & $0.550^{+0.013}_{-0.016}$ & $274^{+22}_{-25}$ & & $-2.483^{+0.020}_{-0.020}$ 
		\\
		\hline
		\multirow{2}{*}{JUNO} & NMO &  \multirow{2}{*}{$0.3092^{+0.0087}_{-0.0087}$}  & \multirow{2}{*}{---} & \multirow{2}{*}{---} & \multirow{2}{*}{---} & \multirow{2}{*}{$7.50^{+0.12}_{-0.12}$} &  \multirow{2}{*}{---}
		\\
		& IMO & & & & & &
		\\
		\hline\hline
	\end{tabular}}
	\caption{Central values of neutrino oscillation parameters together with their $1\sigma$ uncertainties from the NuFIT 6.0/6.1~\cite{Esteban:2024eli,nufit:61} and the latest JUNO results~\cite{Abusleme:2025wem}. $\ell =1 $ is for the normal mass ordering (NMO) with $m^{}_1 < m^{}_2 <m^{}_3$, and $\ell=2$ for the inverted mass ordering (IMO) with $m^{}_3 < m^{}_1 < m^{}_2$.}
	\label{tab:osc-data}
\end{table}

\begin{figure}[t!]
	\centering
	\includegraphics[width=1\linewidth]{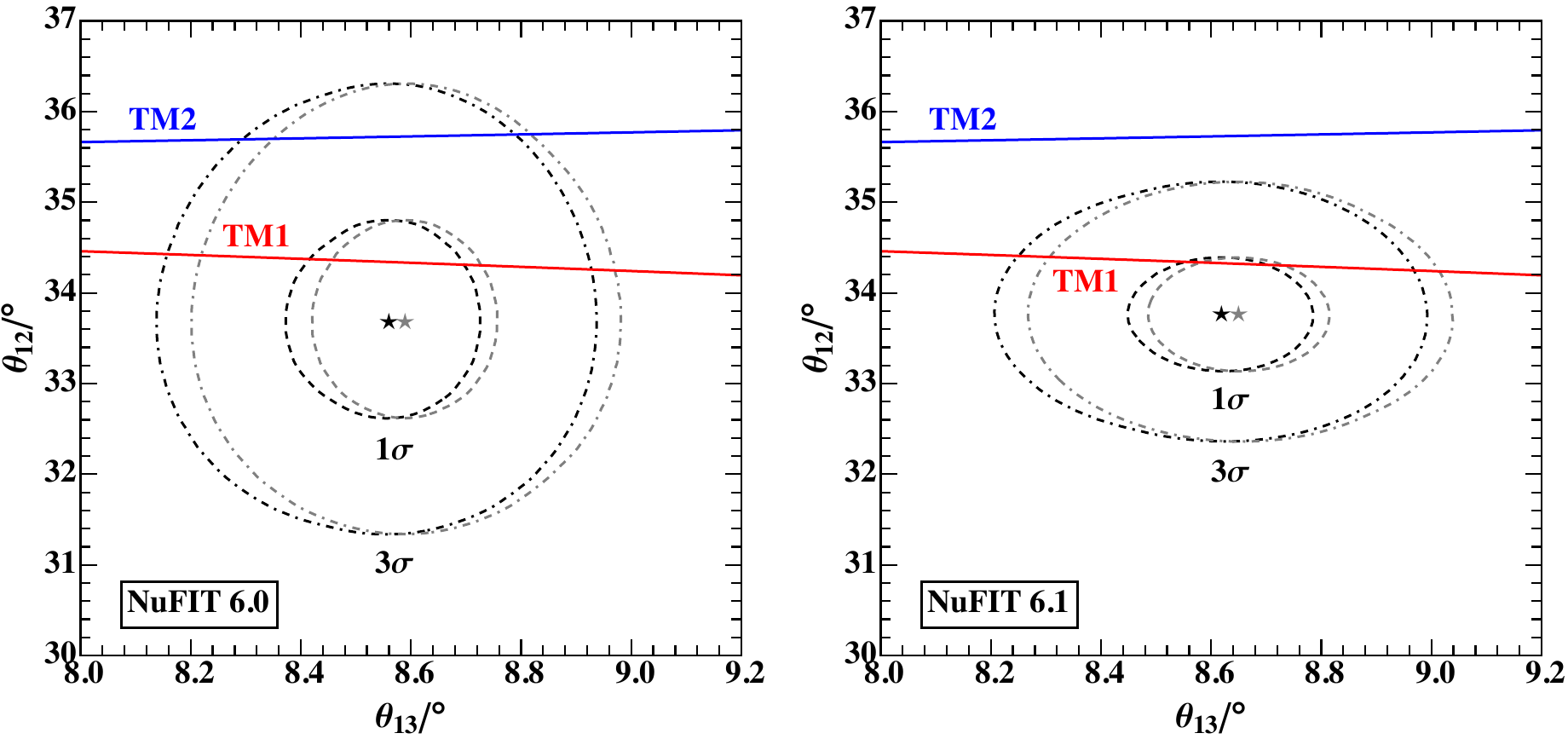}
	\caption{Correlations between $\theta^{}_{12}$ and $\theta^{}_{13}$ in the TM1 and TM2 mixing patterns. The black (grey) stars, dashed lines, and dashed-dotted lines stand for the best-fit point, $1\sigma$ region, and $3\sigma$ region of $\theta^{}_{12}$ and $\theta^{}_{13}$, respectively, in the NMO (IMO) case. In the left panel, the central values of $\theta^{}_{12}$ and $\theta^{}_{13}$ together with their uncertainties are both taken from the NuFIT 6.0 results, while those in the right panel are from the NuFIT 6.1 results incorporating the latest JUNO measurement.}
	\label{fig:correlation}
\end{figure}

Such simple relations between mixing angles are particular striking, as they can be exploited for stringent experimental tests and even potential exclusion of the corresponding mixing patterns and their underlying flavor symmetries, especially now that we are entering the era of subpercent precision in neutrino oscillation measurements~\cite{Capozzi:2025wyn}. The JUNO experiment~\cite{JUNO:2015zny} has very recently released its first measurement based on 59.1 days of data~\cite{Abusleme:2025wem}, and the uncertainties of $\theta^{}_{12}$ and $\Delta m^2_{21} $ have been remarkably reduced with respect to previous measurements. These new measurement results have been incorporated into the latest global analysis of neutrino oscillation data~\cite{Capozzi:2025ovi,nufit:61}, where the JUNO results dominate the global fit for $\theta^{}_{12}$ and $\Delta m^2_{21}$. We summarize the latest JUNO measurement and the NuFIT results from a global three-flavor oscillation analysis in Table~\ref{tab:osc-data}, where the NuFIT 6.1 includes the JUNO measurement while the NuFIT 6.0 does not. For consistency, hereafter we exploit the NuFIT 6.0 and 6.1 results to discuss the implications of the first JUNO measurements. With the most precise measurement of $\theta^{}_{12}$ from the JUNO experiment, the TM mixing patterns are meeting the most serious challenge through the correlations between $\theta^{}_{12}$ and $\theta^{}_{13}$ in Eq.~\eqref{eq:TM-relations}. To clearly see this point, we plot in Fig.~\ref{fig:correlation} the two TM correlations in Eq.~\eqref{eq:TM-relations} together with the best-fit values,  and $1\sigma$ and $3\sigma$ regions of $\theta^{}_{12}$ and $ \theta^{}_{13}$ directly from their two-dimensional $\chi^2$ projections given by the NuFIT 6.0 and 6.1, respectively. It is obvious that with the incorporation of the JUNO result, the experimental data increasingly disfavor the TM mixing patterns. Notably, the TM2 pattern lies substantially outside the $3\sigma$ region, and therefore, it is strongly disfavored in light of the improved precision on $\theta^{}_{12}$, whereas the TM1 pattern sits on the edge of the $1\sigma$ region. However, it should be noted that these flavor mixing patterns are typically realized through certain flavor symmetries at a super-high energy scale~\cite{Grimus:2008tt,deMedeirosVarzielas:2012apl,Luhn:2013lkn,Li:2013jya,Zhao:2014yaa,Girardi:2015rwa,Petcov:2018snn,Novichkov:2018yse,King:2019vhv,Ding:2020vud,Thapa:2021ehj,deMedeirosVarzielas:2021pug,Zhang:2024rwv}. Therefore, when confronting them with low-energy experimental data, one must properly take into account renormalization group (RG) running effects on neutrino parameters (see, e.g., the review~\cite{Ohlsson:2013xva} and references therein). In general, the TM correlations given in Eq.~\eqref{eq:TM-relations} are not invariants against the RG evolution, and consequently, RG corrections to $\theta^{}_{12}$ and $\theta^{}_{13}$ can break their correlations and allow $\theta^{}_{12}$ to deviate from the values predicted by the exact relations. This suggests that the RG evolution of neutrino parameters between energy scales may render the TM mixing patterns compatible with current experimental data, but it critically depends on the direction and magnitude of the running effects. 

In this context, the present work aims to carefully investigate the RG running effects on the TM mixing patterns and to examine whether, or to what extent, these corrections can reconcile the TM predictions with the current and forthcoming precision measurements. We consider both Majorana and Dirac neutrino cases, where the Standard Model is extended with the unique dimension-five Weinberg operator~\cite{Weinberg:1979sa} and with three right-handed neutrinos, respectively, to generate neutrino masses. To remain model-independent and generic, we simply assume that the lepton flavor mixing exhibits the TM mixing patterns at a typical super-high energy scale, without specifying their possible origins from underlying flavor symmetries.

The remainder of this paper is organized as follows. In Section~\ref{sec:TM patterns}, we take a closer look at the TM mixing patterns and briefly discuss their predictions in the absence of RG corrections. The RG equations in the Majorana and Dirac cases are introduced in Section~\ref{sec:RG corrections}, where analytical estimates of the RG running effects on the TM $\theta^{}_{12}$-$\theta^{}_{13}$ correlations are also performed. Section~\ref{sec:Numerical} provides detailed numerical analyses of the RG effects on the TM1 mixing pattern and its compatibility with the latest experimental data, followed by a similar numerical discussion of the TM2 mixing pattern in Section~\ref{sec:TM2}. Finally, we present our summary in Section~\ref{sec:Summary}.

\section{Trimaximal Mixing Patterns}\label{sec:TM patterns}

As shown in the right panel of Fig.~\ref{fig:correlation}, the TM1 mixing pattern still lies within the $3\sigma$ region of $\theta^{}_{12}$ and $\theta^{}_{13}$ when the latest JUNO results are included, whereas the TM2 pattern is already substantially outside the $3\sigma$ region. For this reason, the former appears more appealing and could be more easily brought into better agreement with the experimental data. Therefore, we mainly focus on the TM1 mixing pattern in the following analysis, and then in Section~\ref{sec:TM2}, we will briefly discuss the results for the TM2 mixing pattern.

\begin{figure}[t!]
	\centering
	\includegraphics[width=\linewidth]{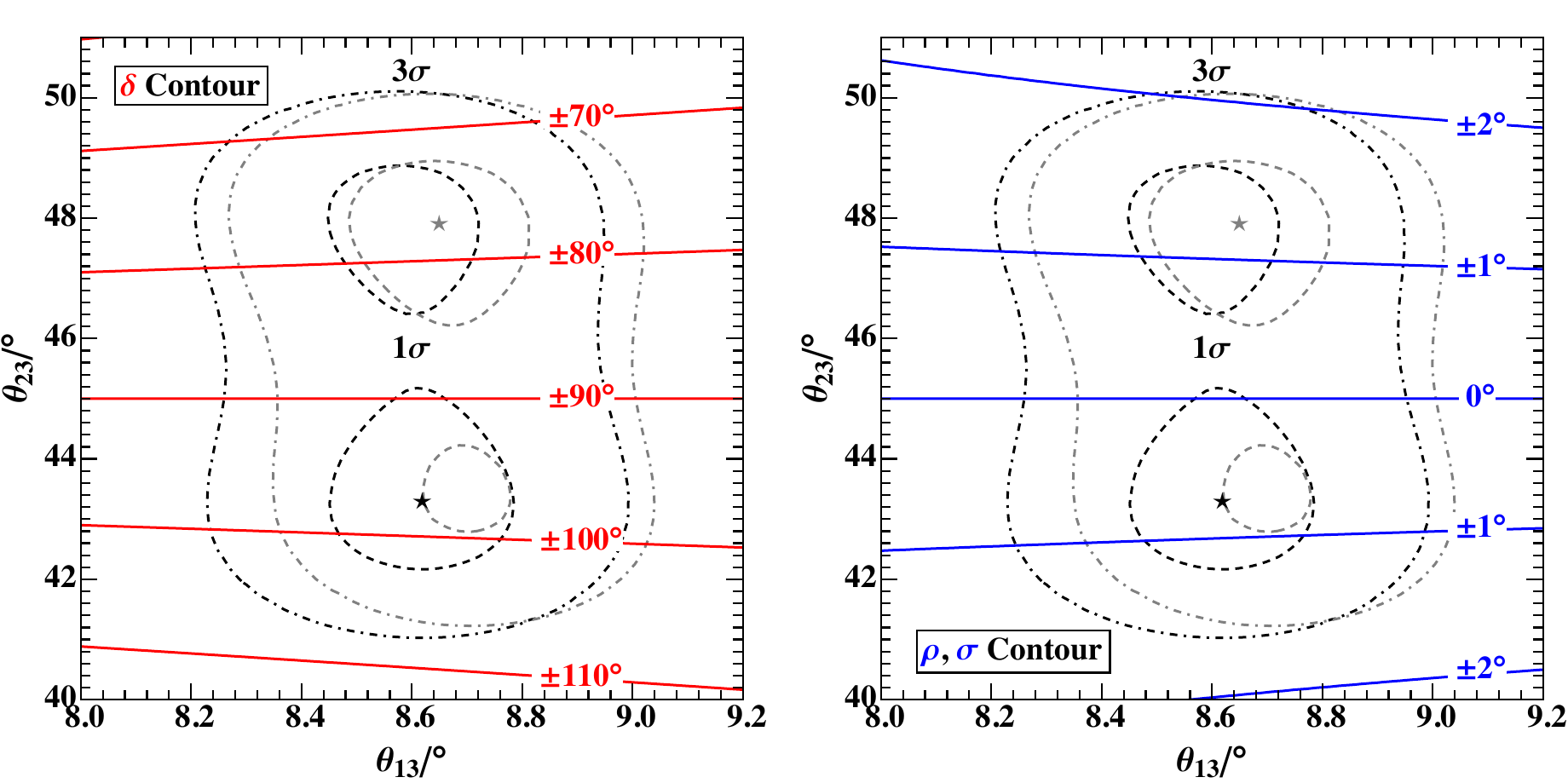}
	%	\begin{subfigure}{\textwidth}
		%	\centering
		%	\includegraphics[width=\linewidth]{fig/del}
		%	\end{subfigure}
	%	\\
	%	\begin{subfigure}{\textwidth}
		%	\centering
		%	\includegraphics[width=\linewidth]{fig/mps}
		%	\end{subfigure}
	\caption{Contour plots for $\delta$ and $\rho \left( \sigma \right)$ in the $\theta^{}_{23}$-$\theta^{}_{13}$ plane in both NMO and IMO cases. The black (grey) stars, dashed lines, and dashed-dotted lines stand for the best-fit point, $1\sigma$ region, and $3\sigma$ region of $\theta^{}_{13}$ and $\theta^{}_{23}$ from the NuFIT 6.1 global analysis.}
	\label{fig:phase}
\end{figure}

Since the TM1 mixing matrix $U^{}_{\rm TM1}$ in Eq.~\eqref{eq:TM1} contains only two free parameters, i.e., the angle $\theta$ and phase $\phi$, which determine all three mixing angles $\theta^{}_{ij}$ (for $ij=12,13,23$), one Dirac CP-violating phase $\delta$, and two Majorana CP-violating phases $\rho$ and $\sigma$ (present only in the Majorana case and unphysical in the Dirac case) under the standard parametrization in Eq.~\eqref{eq:parametrization},  it gives rise to one (three) additional correlation(s) among neutrino mixing parameters in the Dirac (Majorana) case, apart from the correlation between $\theta^{}_{12}$ and $\theta^{}_{13}$ already shown in Eq.~\eqref{eq:TM-relations}, and two relations linking the mixing parameters to the free parameters $\theta$ and $\phi$. These relations are found to be
\begin{eqnarray} \label{eq:TM1-relations}
	&&\sin\theta^{}_{13} = \frac{1}{\sqrt{3}} \sin\theta \;,\quad \cos2\theta^{}_{23} = -2\sqrt{2} \tan\theta^{}_{13} \sqrt{1-2\tan^2\theta^{}_{13}} \cos\phi \;, \quad \sin\phi=\sin2\theta^{}_{23}\sin\delta \;,
	\nonumber
	\\
	&&\cos\delta = - \frac{\cot2\theta^{}_{23} \left( 1 - 5\sin^2\theta^{}_{13} \right)}{2\sqrt{2} \sin\theta^{}_{13} \sqrt{1-3\sin^2\theta^{}_{13}}} \;,\quad \rho=\sigma=\phi-\delta \;.
\end{eqnarray}
The last relation for $\rho$ and $\sigma$ in the above equation is physically meaningful only in the Majorana case, and we keep $\phi$ and $\delta$ explicitly in the relation, instead of substituting their expressions in terms of $\theta^{}_{13}$ and $\theta^{}_{23}$, for brevity. The three relations in the first row of Eq.~\eqref{eq:TM1-relations} establish connections between the mixing parameters and the two free parameters, and it is convenient to use the first two of them as conversion relations between the $(\theta, \phi)$ coordinate and the $(\theta^{}_{13}, \theta^{}_{23})$ coordinate systems. Consequently, all other mixing parameters can be determined from a given pair of $(\theta^{}_{13}, \theta^{}_{23})$ values through the correlations in Eq.~\eqref{eq:TM-relations} and in the last line of Eq.~\eqref{eq:TM1-relations}. The dependence of $\theta^{}_{12}$ on $\theta^{}_{13}$ has been presented in Fig.~\ref{fig:correlation}, and here, we show contour plots for $\delta$ and $\rho \left( \sigma \right)$ in the $\theta^{}_{23}$-$\theta^{}_{13}$ plane in Fig.~\ref{fig:phase}, where the best-fit values, 1$\sigma$ and 3$\sigma$ regions of $\left( \theta^{}_{13}, \theta^{}_{23} \right)$ from the NuFIT 6.1 are also shown.

As seen from the left panel of Fig.~\ref{fig:phase}, each contour line of $\delta$ corresponds to two values with opposite signs due to the ambiguity in $\cos\delta$, and the quadrant of $\delta$ is fully correlated with the octant of $\theta^{}_{23}$. More specifically, the first (second) octant of $\theta^{}_{23}$ results in the second or third (first or fourth) quadrant of $\delta$. As the best-fit values of $\theta^{}_{23}$ in the NMO and IMO cases are in different octants, different quadrants of $\delta$ are preferred in these two cases.  Inputting the best-fit values of $\theta^{}_{13}$ and $\theta^{}_{23}$ from the NuFIT 6.1 listed in Table~\ref{tab:osc-data}, one gets $\delta = 97.49^\circ$ or $262.51^\circ$ in the NMO case while $\delta = 77.47^\circ$ or $282.53^\circ$ in the IMO case. The latter values in the two cases lie within the current experimental $3\sigma$ and $1\sigma$ ranges of $\delta$, respectively. Note that $\theta^{}_{23} = 45^\circ$ leads to $\delta = \phi = \pm 90^\circ$, and then the TM1 mixing matrix possesses a $\mu$-$\tau$ symmetry, namely $|U^{}_{\mu i}| = |U^{}_{\tau i}|$ (for $i=1,2,3$)~\cite{Xing:2014zka}. Since the present best-fit value of $\theta^{}_{23}$ slightly deviates from $45^\circ$, the TM1 mixing pattern with an approximate $\mu$-$\tau$ symmetry is favored by the current experimental data. As shown in the right panel of Fig.~\ref{fig:phase}, the contour lines of $\rho$ and $\sigma$ are symmetric with respect to $\rho, \sigma = 0 $, corresponding to $\theta^{}_{23} = 45^\circ$, and each line also represents two values with opposite signs. Moreover, within the experimental $3\sigma $ ranges of $\theta^{}_{23}$ and $\theta^{}_{13}$, the values of $\rho$ and $\sigma$ are very tiny. In fact, $\rho$ and $\sigma$ capture the breaking of the $\mu$-$\tau$ symmetry, and hence their tiny values again indicate that the current experimental date are in favor of an approximate $\mu$-$\tau$ symmetry embedded in the TM1 mixing pattern. One can take advantage of the relations in Eq.~\eqref{eq:TM1-relations} to understand the aforementioned features of $\rho$ and $\sigma$. Taking into account the experimentally allowed ranges of $\theta^{}_{23}$ and $\theta^{}_{13}$, the relations in Eq.~\eqref{eq:TM1-relations} infer that $\cos\delta/\cos\phi>0$ and $\sin\delta/\sin\phi >0$ hold, and thereby, $\delta$ and $\phi$ are always in the same quadrant. As the difference between $\phi$ and $\delta$, the signs and magnitudes of $\rho$ and $\sigma$ are governed by the relation $\sin\phi = \sin2\theta^{}_{23} \sin\delta$, which is symmetric with respect to $\theta^{}_{23} =45^\circ$. The fact that the experimental measurement of $\theta^{}_{23}$ is very close to $45^\circ$ leads to small values of $\rho=\sigma = \phi-\delta$. In turn, the smallness of $\rho$ and $\sigma$ quantifies the deviation of $\theta^{}_{23}$ from $45^\circ$, or equivalently, the extent of $\mu$-$\tau$ symmetry breaking. Because of the ambiguous quadrants of $\phi$ and $\delta$, the signs of $\rho$ and $\delta$ remain undetermined: they are positive when $\phi$ and $\delta$ lie in the second or fourth quadrant, and negative when $\phi$ and $\delta$ lie in the first or third quadrant.

\section{Renormalization Group Running Effects} \label{sec:RG corrections}

Since the CP-violating phase(s) have not yet been well measured, and the Majorana or Dirac nature of neutrinos even remains an open question, the $\theta^{}_{12}$-$\theta^{}_{13}$ correlations in Eq.~\eqref{eq:TM-relations}, being independent of the CP-violating phase(s), can serve as a distinctive feature of the TM mixing patterns and be exploited to test them, particularly now that the latest JUNO results have significantly improved the precision of $\theta^{}_{12}$. As clearly shown in Fig.~\ref{fig:correlation}, the $\theta^{}_{12}$-$\theta^{}_{13}$ correlations predicted by the TM1 and TM2 mixing patterns already lie outside the experimentally allowed $1\sigma$ and $3\sigma$ region, respectively, after taking into account the latest JUNO results. However, it is necessary to properly include RG corrections to these relations when confronting them with low-energy measurements, as the TM mixing patterns are typically realized through underlying flavor symmetries at a super-high energy scale. In this section, we introduce the RG equations for the neutrino mass matrix in both Majorana and Dirac neutrino cases and examine whether RG corrections can bring the $\theta^{}_{12}$-$\theta^{}_{13}$ correlations into better agreement with current experimental measurements. Here we do not discuss the underlying origin of the TM patterns, but instead simply assume that they are realized at a super-high energy scale, i.e., $\Lambda = 10^{16}$ GeV.

Considering that neutrino masses in the Majorana and Dirac cases are generated through the Weinberg operator~\cite{Weinberg:1979sa} and the Higgs mechanism~\cite{Higgs:1964ia,Englert:1964et,Higgs:1964pj,Guralnik:1964eu}, respectively, the one-loop RGE for the neutrino mass matrix $M^{}_\nu$ can be written as~\footnote{The two-loop RGE of the Weinberg operators has been derived in Refs.~\cite{Davidson:2006tg,Ibarra:2024tpt}. It is unnecessary to include it in the present case due to its relative smallness compared with the one-loop effects. However, in the case where the lightest neutrino is massless, these two-loop RG effects can generate a non-zero mass for this initially massless neutrino~\cite{Davidson:2006tg,Ibarra:2024tpt,Xing:2020ezi} while the one-loop effects can not do this job~\cite{Ibarra:2018dib,Ibarra:2020eia,Zhang:2024weq}.}
\begin{eqnarray}\label{eq:rge-m}
	16\pi^2 \mu \frac{{\rm d} M^{}_\nu}{{\rm d} \mu} = \alpha^{}_M M^{}_\nu - \frac{3}{2} \left( Y^{}_l Y^\dagger_l \right) M^{}_\nu - \frac{3}{2} M^{}_\nu \left( Y^{}_l Y^\dagger_l \right)^{\rm T}
\end{eqnarray}
in the Majorana case~\cite{Chankowski:1993tx,Babu:1993qv,Antusch:2001ck}, and 
\begin{eqnarray}\label{eq:rge-d}
	16\pi^2 \mu \frac{{\rm d} M^{}_\nu}{{\rm d} \mu} = \alpha^{}_D M^{}_\nu - \frac{3}{2} \left( Y^{}_l Y^\dagger_l -Y^{}_\nu Y^\dagger_\nu \right) M^{}_\nu 
\end{eqnarray}
in the Dirac case~\cite{Cheng:1973nv,Machacek:1983fi,Grzadkowski:1987tf}, where the flavor-blind factors $\alpha^{}_M$ and $\alpha^{}_D$ is given by
\begin{eqnarray}
	\alpha^{}_M &=& 4\lambda - 3g^2_2 + 2 {\rm Tr} \left( Y^{}_l Y^\dagger_l + 3 Y^{}_{\rm u} Y^\dagger_{\rm u} +  3Y^{}_{\rm d} Y^\dagger_{\rm d} \right) \;,
	\nonumber
	\\
	\alpha^{}_D &=& -\frac{3}{4}\left( g^2_1 + 3g^2_2 \right) +  {\rm Tr} \left( Y^{}_\nu Y^\dagger_\nu + Y^{}_l Y^\dagger_l + 3 Y^{}_{\rm u} Y^\dagger_{\rm u} +  3Y^{}_{\rm d} Y^\dagger_{\rm d} \right) \;,
\end{eqnarray}
and $\mu$ is the renormalization energy scale. Given the fact that $Y^{}_\nu/Y^{}_l \lesssim 10^{-6}$ holds in the Dirac case, one can safely omit $Y_\nu$ in Eq.~\eqref{eq:rge-d} as well as those in $\alpha^{}_D$ and other couplings' RGEs. This leaves all couplings' RGEs the same in both two cases, except for the ones of $M^{}_\nu$. Working in the basis where $Y^{}_l $ is diagonal, i.e., $Y^{}_l = {\rm Diag} \{ y^{}_e, y^{}_\mu, y^{}_\tau \}$, at the initial energy scale $\Lambda$, $Y^{}_l$ remains diagonal during RG running. In this case, the lepton flavor mixing always arises entirely from the diagonalization of the neutrino mass matrix, namely, $U^\dagger M^{}_\nu U^\ast = {\rm Diag} \{ m^{}_1, m^{}_2, m^{}_3\}$ or $U^\dagger H^{}_\nu U = {\rm Diag} \{ m^2_1, m^2_2, m^2_3\}$  in the Majorana or Dirac case, where $m^{}_i$ (for $i=1,2,3$) denote neutrino masses, and $H^{}_\nu \equiv M^{}_\nu M^\dagger_\nu $ is introduced in the Dirac case. Furthermore, the integral solution to the neutrino mass matrix's RGE turns out to be~\cite{Zhang:2020lsd}
\begin{eqnarray}\label{eq:integral-M}
	M^{}_\nu \left( \mu \right) = I^{}_M  T^{}_l M^{}_\nu \left( \Lambda \right) T^{}_l \;,
\end{eqnarray}
in the Majorana case, and
\begin{eqnarray}\label{eq:integral-D}
	H^{}_\nu \left( \mu \right) = I^2_D T^{}_l H^{}_\nu \left( \Lambda \right) T^{}_l \;,
\end{eqnarray}
in the  Dirac case, where $T^{}_l = {\rm Diag} \{ I^{}_e, I^{}_\mu, I^{}_\tau \}$ is the same in two cases, and the scale-dependent running factors $I^{}_{M,D}$ and $I^{}_{\alpha}$ (for $\alpha=e,\mu,\tau$) are given by
\begin{eqnarray}
	I^{}_{M,D} \left( \mu \right) &=& \exp \left[ \frac{1}{16\pi^2} \int^{\ln\mu}_{\ln\Lambda} \alpha^{}_{M,D} \; {\rm d} \ln\mu^\prime \right] \;,
	\nonumber
	\\
	I^{}_\alpha  \left( \mu \right)  &=& \exp\left[ -\frac{3}{32\pi^2} \int^{\ln\mu}_{\ln\Lambda} y^2_\alpha \; {\rm d} \ln\mu^\prime \right] \;.
\end{eqnarray}
Obviously, $I^{}_{M}$ and $I^{}_{D}$ are flavor-blind, and they only affect the magnitude of neutrino masses universally, whereas $I^{}_\alpha$ (for $\alpha=e,\mu,\tau$) can change the flavor structure of $M^{}_\nu$ or $H^{}_\nu$ and hence dominate the running of flavor mixing parameters. Considering $y_e \ll y^{}_\mu \ll y^{}_\tau \ll 1$, we have $T^{}_l \simeq {\rm Diag} \{ 1, 1, 1+\Delta^{}_\tau \}$ with 
\begin{eqnarray}\label{eq:deltatau}
	\Delta^{}_\tau \left( \mu \right) = -\frac{3}{32\pi^2} \int^{\ln\mu}_{\ln\Lambda} y^2_\tau \; {\rm d} \ln\mu^\prime \;.
\end{eqnarray}
Fig.~\ref{fig:runningfactors} shows the values of running factors $I^{}_{M,D}$ and $\Delta^{}_\tau$ against the energy scale $\mu$ from $\Lambda=10^{16}$~GeV down to the electroweak scale $\Lambda^{}_{\rm EW}  = 200$ GeV. It manifests that $\Delta^{}_\tau$ gains a very tiny value around $3\times 10^{-5}$ at $\Lambda^{}_{\rm EW}  $ and consequently, RG corrections to mixing angles and phases are in general very small. Fortunately, RG corrections to the mixing angle $\theta^{}_{12}$ can be significantly enhanced by the factor $\left( m^{}_2 + m^{}_1 \right)/\left( m^{}_2 - m^{}_1 \right)$ or $\left( m^2_2 + m^2_1 \right)/\left( m^2_2 - m^2_1 \right)$ when neutrinos have a nearly degenerate mass spectrum~\cite{Ohlsson:2013xva,Zhang:2020lsd,Casas:1999tp,Casas:1999ac,Casas:1999tg,Antusch:2003kp,Antusch:2005gp,Mei:2005qp}. Such an enhancement of $\theta^{}_{12}$ is highly desirable in the present context, and moreover, a nearly degenerate mass spectrum would be sensitivity to neutrino non-oscillation experiments, such as beta decay and neutrinoless double beta decay experiments. However, in addition to the size of RG corrections, an equally crucial factor is their direction, which determines whether these corrections drive $\theta^{}_{12}$ toward or away from its experimentally measured value. To demonstrate these essential points, we reproduce here the general expression of the RG correction to $\theta^{}_{12}$, i.e., $\Delta \theta^{}_{12} \left( \mu \right) \equiv \theta^{}_{12} \left( \mu\right) - \theta^{}_{12} \left( \Lambda \right)$, which can be approximately derived from Eq.~\eqref{eq:integral-M} or \eqref{eq:integral-D}~\cite{Zhang:2020lsd}:
\begin{figure}[t!]
	\centering
	\includegraphics[width=1\linewidth]{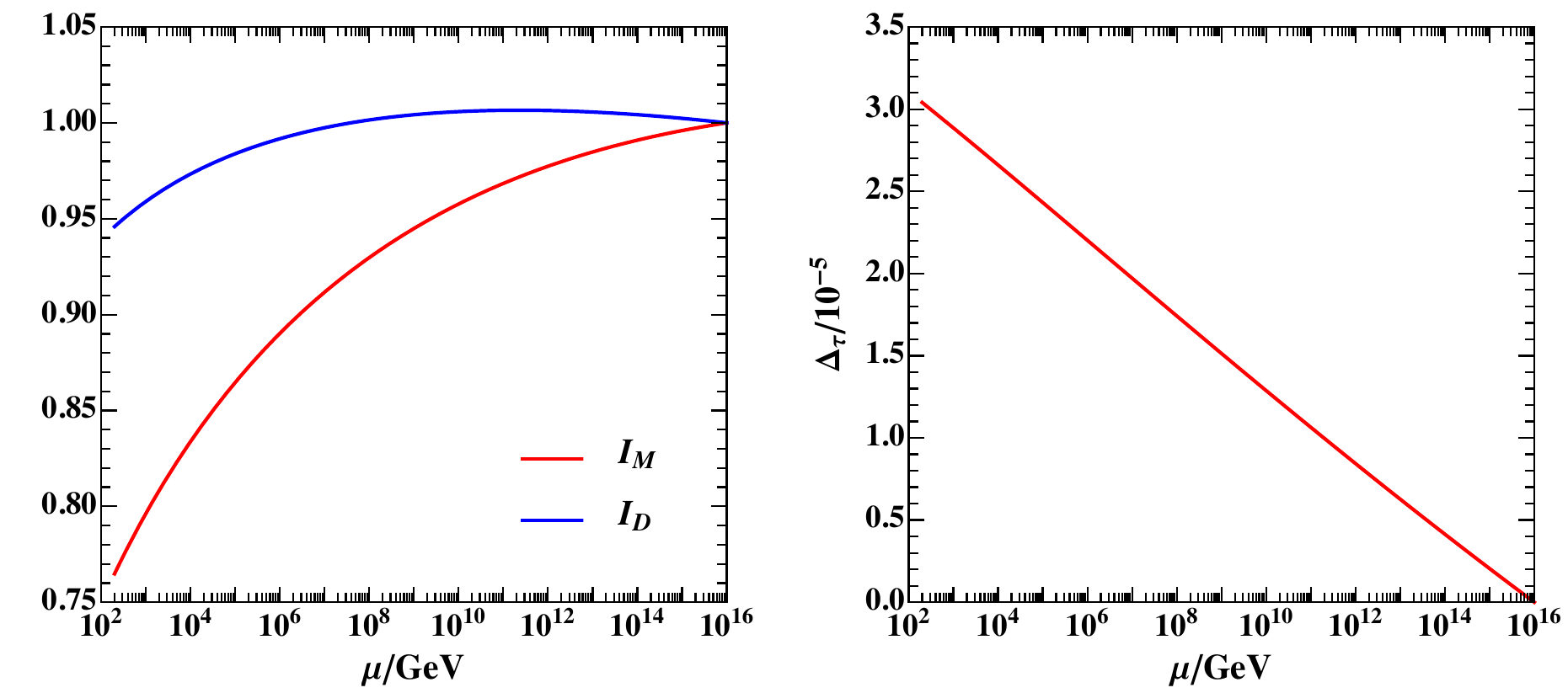}\vspace{-0.cm}
	\caption{The values of running factors $I^{}_{M,D}$ and $\Delta^{}_\tau$ against the energy scale $\mu$ from $\Lambda=10^{16}$~GeV down to the electroweak scale $\Lambda^{}_{\rm EW}  = 200$ GeV.}
	\label{fig:runningfactors}
\end{figure}
\begin{eqnarray}\label{eq:rge-th12-M}
	\Delta \theta^{}_{12}  &\simeq&   \frac{\Delta^{}_\tau}{2} \left\{ \vphantom{\frac{1}{1}} \sin{2\theta^{}_{12}} \sin^2{\theta^{}_{13}} \cos^2{\theta^{}_{23}} \left[ \zeta^{}_{32} \sin^2{ \left( \delta + \sigma \right) } - \zeta^{}_{31} \sin^2{ \left( \delta + \rho \right)}  + \zeta^{-1}_{32} \cos^2{\left( \delta + \sigma \right)} \right.\right.
	\nonumber
	\\
	&& - \left. \zeta^{-1}_{31} \cos^2{\left( \delta + \rho \right)} \right] - \left[ \left( \sin^2{\theta^{}_{23}} - \sin^2{\theta^{}_{13}} \cos^2{\theta^{}_{23}} \right) \sin{2\theta^{}_{12}} - \sin{\theta^{}_{13}}\sin{2\theta^{}_{23}} \cos{2\theta^{}_{12}} \right.
	\nonumber
	\\
	&& \times \left. \cos{\delta} \vphantom{\cos^2} \right] \left[ \zeta^{}_{21} \sin^2{\left( \rho - \sigma\right)} + \zeta^{-1}_{21}\cos^2{\left(\rho-\sigma\right)} \right]  + \sin{2\theta^{}_{23}} \sin{\theta^{}_{13}} \left[ \left( \zeta^{}_{31} -\zeta^{-1}_{31} \right)\sin^2{\theta^{}_{12}} \right.
	\nonumber
	\\
	&& \times \left. \sin{(\delta+\rho)}\sin{\rho} + \left( \zeta^{}_{32} -\zeta^{-1}_{32} \right)\cos^2{\theta^{}_{12}}\sin{(\delta+\sigma)}\sin{\sigma} + \frac{1}{2} \left( \zeta^{}_{21} -\zeta^{-1}_{21} \right) \sin{2\left( \rho - \sigma\right)} \right.
	\nonumber
	\\
	&& \times \left.\left. \sin{\delta} + \left( \zeta^{-1}_{31} \sin^2{\theta^{}_{12}} + \zeta^{-1}_{32} \cos^2{\theta^{}_{12}} \right)\cos{\delta}  \right] \vphantom{\frac{1}{1}} \right\} \;,
\end{eqnarray}
with $\zeta^{}_{ij} = \left( m^{}_i - m^{}_j \right)/ \left( m^{}_i + m^{}_j \right) $ in the Majorana case, and 
\begin{eqnarray}\label{eq:rge-th12-D}
	\Delta \theta^{}_{12} &\simeq& \frac{\Delta^{}_\tau}{2} \left[ \sin\theta^{}_{13} \sin{2\theta^{}_{23}} \cos\delta \left( \xi^{}_{21} \cos2\theta^{}_{12} +  \xi^{}_{31} \sin^2\theta^{}_{12} + \xi^{}_{32} \cos^2\theta^{}_{12} \right) \right.
	\nonumber
	\\
	&& \left. + \sin2\theta^{}_{12} \sin^2\theta^{}_{13} \cos^2\theta^{}_{23} \left( \xi^{}_{21} + \xi^{}_{32} - \xi^{}_{31} \right) - \xi^{}_{21} \sin2\theta^{}_{12} \sin^2\theta^{}_{23} \right] \;,
\end{eqnarray}
with $\xi^{}_{ij} = \left( m^2_i + m^2_j \right) / \left( m^2_i - m^2_j \right)$ in the Dirac case. The approximate integral results for neutrino masses and other mixing parameters in both cases can be found in Ref.~\cite{Zhang:2020lsd}. 

Now, let us define a function that quantifies the deviation from the TM1 $\theta^{}_{12}$-$\theta^{}_{13}$ correlation in  Eq.~\eqref{eq:TM-relations} at the energy scale $\mu$:
\begin{eqnarray}\label{eq:fun}
	f\left(\mu\right) = \sin^2 \theta^{}_{12} \left( \mu \right)  + \frac{2}{3} \tan^2\theta^{}_{13} \left( \mu \right) - \frac{1}{3} \;.
\end{eqnarray}
One immediately has $f\left( \Lambda \right) = 0$ as indicated by the exact $\theta^{}_{12}$-$\theta^{}_{13}$ correlation, and at the leading order, $f\left( \mu\right)$ can be expanded as
\begin{eqnarray}\label{eq:fun-rge}
	f\left( \mu \right) &\simeq& \sin2\theta^{}_{12} \left( \Lambda \right) \cdot \Delta \theta^{}_{12} \left( \mu \right) + \frac{4}{3} \tan\theta^{}_{13} \left( \Lambda \right)  \sec^2\theta^{}_{13} \left( \Lambda \right) \cdot \Delta\theta^{}_{13} \left( \mu \right)
	\nonumber
	\\
	&\simeq& \sin2\theta^{}_{12} \left( \Lambda^{}_{\rm EW} \right) \cdot \Delta \theta^{}_{12} \left( \mu \right) + \frac{4}{3} \tan\theta^{}_{13} \left( \Lambda^{}_{\rm EW} \right)  \sec^2\theta^{}_{13} \left( \Lambda^{}_{\rm EW} \right) \cdot \Delta\theta^{}_{13} \left( \mu \right) \;,
\end{eqnarray}
where $f\left( \Lambda \right) = 0$ has been used. Generally, $f\left( \mu \right) = 0$ is not protected against RG corrections, and any deviation of $f\left( \mu \right)$ from 0 stands for a breaking of the TM1 mixing pattern. If one simply substitutes the latest best-fit values for $\theta^{}_{12}= 33.76^\circ$ and $\theta^{}_{13} = 8.62^\circ$ from NuFIT 6.1 into Eq.~\eqref{eq:fun},  it gives $f^{}_{\rm exp} \left( \Lambda^{}_{\rm EW} \right) \simeq -9\times 10^{-3} $. Therefore, if RG effects could sizeably decrease $f\left( \mu \right)$ from $f\left( \Lambda \right) = 0$ to $f^{}_{\rm exp} \left( \Lambda^{}_{\rm EW} \right) \simeq -9\times 10^{-3} $, the TM1 $\theta^{}_{12}$-$\theta^{}_{13}$ correlation incorporating RG corrections would become fully consistent with the latest experimental data. Considering that the factor $\Delta^{}_\tau \left( \Lambda^{}_{\rm EW } \right) \simeq 3\times 10^{-5}$ governing the overall size of RG corrections to mixing angles is much smaller than $9\times10^{-3}$, the only possibility of achieving such a large correction in the present context is that neutrino masses are highly degenerate. In this case, $\Delta \theta^{}_{12}$ is largely enhanced by $\zeta^{-1}_{12}$ or $\xi^{}_{12}$ while $\Delta \theta^{}_{13}$ is not, and hence the second term involving $\Delta \theta^{}_{13}$ in Eq.~\eqref{eq:fun-rge} can be simply omitted. With the help of Eq.~\eqref{eq:rge-th12-M} or \eqref{eq:rge-th12-D} and taking into account relations in Eqs.~\eqref{eq:TM-relations} and \eqref{eq:TM1-relations}, $f\left( \Lambda^{}_{\rm EW} \right)$ in Eq.~\eqref{eq:fun-rge} is approximately given by
\begin{eqnarray}\label{eq:fun-app}
	f \left( \Lambda^{}_{\rm EW} \right) &\sim& \left\{ \begin{matrix}- \dfrac{4\Delta^{}_\tau}{9} \zeta^{-1}_{21}  \sin^2{\theta^{}_{23}}\;, & \qquad{\rm (Majorana~case)} \\[0.35cm] - \dfrac{4\Delta^{}_\tau}{9} \xi^{}_{21} \sin^2\theta^{}_{23} \;, & \qquad{\rm (Dirac~case)}  \end{matrix}  \right. \;.
\end{eqnarray}
Fortunately, it turns out that the RG corrections in the both Majorana and Dirac cases are in the right direction that leads to a negative value of $f \left( \Lambda^{}_{\rm EW} \right) $. Now, requiring $f \left( \Lambda^{}_{\rm EW} \right) = f^{}_{\rm exp} \left( \Lambda^{}_{\rm EW} \right) \simeq -9\times 10^{-3}$ and inputting the best-fit values of $\theta^{}_{23} = 43.29^\circ$ and $\Delta m^2_{21} = 7.537\times 10^{-5}~{\rm eV}^2$, one obtains $m^{}_1 \sim 0.16$~eV in the Majorana case or $m^{}_1 \sim 0.23$~eV in the Dirac case. The latter is twice larger than the former due to $\zeta^{-1}_{12} \sim 2 \xi^{}_{21}$ in the almost degenerate case. Therefore, in the both Majorana and Dirac cases, the RG corrections with nearly degenerate neutrino masses act in the right direction and are sufficiently large to reconcile the TM1 mixing pattern with experimental data, including the latest JUNO results. However, a nearly degenerate mass spectrum in the Majorana case may encounter severe tension with current neutrinoless double beta decay experiments while the Dirac case remain unaffected. We will discuss these issues in detail when performing more precise and comprehensive numerical analyses in the next section.

\section{Numerical Analyses}\label{sec:Numerical}

In the previous section, we have analytically estimated the RG effects on the TM1 $\theta^{}_{12}$-$\theta^{}_{13}$ correlation in both Majorana and Dirac cases, and found that the TM1 mixing pattern can become fully consistent with the latest JUNO result once RG corrections are included. In this section, we scrutinize these RG effects on the TM1 mixing pattern by performing comprehensive numerical analyses and discuss some constraints from neutrino non-oscillation experiments.

We adopt the $\chi^2$ technique to quantify how well the TM1 mixing pattern incorporating RG effects agrees with the current experimental data and to extract the allowed parameter space. Given the distinct experimental sensitivities, we construct the total $\chi^2$ function as the sum of individual contributions from the solar, reactor, and atmospheric sectors, namely,
\begin{eqnarray}\label{eq:chi-fun}
	\chi^2 = \chi^2_{\rm Solar} \left[ \theta^{}_{12} \left( \bm{p} \right), \Delta m^2_{21} \left( \bm{p} \right) \right] + \chi^2_{\rm Atmos} \left[ \theta^{}_{23} \left( \bm{p} \right), \Delta m^2_{3\ell} \left( \bm{p} \right) \right] + \chi^2_{\rm Reactor} \left[ \theta^{}_{13} \left( \bm{p} \right) \right] \;,
\end{eqnarray} 
where the small correlations among these sectors are neglected, and the $\chi^2_i$ (for $i=$ Solar, Atmos, and Reactor) in each sector is taken from the projected $\chi^2$ data provided by NuFIT 6.1~\cite{nufit:61}~\footnote{The mass ordering preference is removed from the $\chi^{}_i$ data provided by NuFIT 6.1~\cite{nufit:61}. In other words, we subtract $\Delta \chi^2_{\rm MO} = 5.9$ from each $\chi^{}_i$ function in the inverted mass ordering.}. Each $\chi^2_i$ and the total $\chi^2$ are functions of the initial inputs $\bm{p} = \{ m^{}_{\rm lightest}, \Delta m^2_{21}, \Delta m^2_{3\ell}, \theta, \phi \}$ at $\Lambda = 10^{16}$~GeV with $m^{}_{\rm lightest} = m^{}_1$, $\ell = 1$ for the NMO and $m^{}_{\rm lightest} = m^{}_3$, $\ell = 2$ for the IMO. We exclude the Dirac CP-violating phase $\delta$ from the definition of $\chi^2$-function,  since its value is still subject to a large uncertainty and even shows a mild tension between the T2K and NO$\nu$A measurements in the NMO case~\cite{Esteban:2024eli,NOvA:2021nfi,T2K:2023smv,T2K:2025wet}. We take advantage of the gradient descent method~\cite{Ruder:2016lil} to minimize the $\chi^2$ function and obtain the corresponding best-fit initial inputs for $\{ m^{}_{\rm lightest}, \Delta m^2_{21}, \Delta m^2_{3\ell}, \theta, \phi  \}$. Starting with this set of best-fit initial inputs, we then employ the Metropolis algorithm~\cite{Metropolis:1953} to explore the allowed parameter space. 

Besides neutrino masses, mixing angles, and CP-violating phase(s), we also calculate two non-oscillation observables, namely, the two effective neutrino masses, $m^{}_\beta = \sqrt{\sum^{3}_{i=1} m^2_i |U^{}_{ei}|^2}$ and $m^{}_{\beta\beta} = \left|\sum^{3}_{i=1} m^{}_i U^2_{ei} \right|$, which can be probed in beta decay and neutrinoless double beta decay experiments, respectively. The most stringent current constraints on $m^{}_\beta$ and $m^{}_{\beta\beta}$ are provided by KATRIN and KamLAND-Zen experiments, which set upper limits of $m^{}_\beta < 0.45$~eV~\cite{KATRIN:2024cdt} and $m^{}_{\beta\beta} < 0.028-0.122$~eV~\cite{KamLAND-Zen:2024eml}, respectively. These upper limits may impose strong constraints on the allowed parameter space, as a sizable lightest neutrino mass is required to enhance the RG effects. However, the constraint from neutrinoless double beta decay experiments is relevant only in the Majorana case, and it is absent in the Dirac case.

\subsection{The Majorana Case}

\begin{table}[t!]
	\centering
	\renewcommand{\arraystretch}{1.2}
	\begin{tabular}{c|c|c|c|c|c}
		\hline\hline
		& & \multicolumn{2}{c|}{The Majorana Case} &  \multicolumn{2}{c}{The Dirac Case}
		\\
		\cline{3-6}
		& & NMO & IMO & NMO & IMO
		\\
		\hline
		\multirow{5}{*}{\rotatebox{90}{Inputs at $\Lambda$}} & $m^{}_{\rm lightest}/{\rm eV}$ & 0.21703 & 0.19384 & 0.24893 & 0.23943
		\\
		%\cline{2-6}
		& $\Delta m^2_{21}/10^{-5}{~\rm eV^2}$ & 12.790 & 12.761 & 8.3615 & 8.3377
		\\
		%\cline{2-6}
		& $\Delta m^2_{3\ell}/10^{-3}{~\rm eV^2}$ & 4.2895 & 4.2475 & 2.8047 & 2.7767
		\\
		%\cline{2-6}
		& $\theta/^\circ$ &15.059  & 15.093 & 15.069 & 15.084
		\\
		%\cline{2-6}
		& $\phi~({\rm or}~360^\circ-\phi)/^\circ$ & 98.033 & 76.387 & 98.033 & 76.398
		\\
		\hline
		\multirow{11}{*}{\rotatebox{90}{Predictions at $\Lambda^{}_{\rm EW}$}} & $m^{}_{\rm lightest}/{\rm eV}$ & 0.16599 & 0.14826 & 0.23547 & 0.22648
		\\
		%\cline{2-6}
		& $\Delta m^2_{21}/10^{-5}{~\rm eV^2}$  & 7.5334 & 7.5336 & 7.5336 & 7.5361
		\\
		%\cline{2-6}
		& $\Delta m^2_{3\ell}/10^{-3}{~\rm eV^2}$ & 2.5105 & $- 2.4846$ & 2.5107 & $-2.4843$
		\\
		%\cline{2-6}
		& $\theta^{}_{12}/^\circ$ & 33.770 & 33.765 & 33.766 & 33.710
		\\
		%\cline{2-6}
		& $\theta^{}_{13}/^\circ$ & 8.6268 & 8.6464 & 8.6268 & 8.6463
		\\
		%\cline{2-6}
		& $\theta^{}_{23}/^\circ$ & 43.281 & 47.869 & 43.280 & 47.870
		\\
		%\cline{2-6}
		& $\delta~({\rm or}~360^\circ-\delta)/^\circ$ & 97.295 & 77.605 & 96.889 & 77.230
		\\
		%\cline{2-6}
		& $\rho~({\rm or}~180^\circ-\rho)/^\circ$ & 0.74035 & 178.78 & --- & ---
		\\
		%\cline{2-6}
		& $\sigma~({\rm or}~180^\circ-\rho)/^\circ$ & 0.72729 & 178.79 & --- & ---
		\\
		%\cline{2-6}
		& $m^{}_\beta/{\rm eV}$ & 0.16623 & 0.15607 & 0.23563 & 0.23167
		\\
		%\cline{2-6}
		& $m^{}_{\beta\beta}/{\rm eV}$ & 0.15858 & 0.14974 & --- & ---
		\\
		\hline
		& $\chi^2_{\rm min}$ & $1.0 \times 10^{-2}$ & $1.0\times 10^{-2}$ & $1.0\times 10^{-2}$ & $5.9\times 10^{-3}$
		\\
		\hline\hline
	\end{tabular}
	\caption{Initial inputs for $\{ m^{}_{\rm lightest}, \Delta m^2_{21}, \Delta m^2_{3\ell}, \theta, \phi  \}$ in the TM1 mixing pattern at $\Lambda$ that minimize $\chi^2$, and  predictions for $\{ m^{}_{\rm lightest}, \Delta m^2_{21}, \Delta m^2_{3\ell}, \theta^{}_{12}, \theta^{}_{13},\theta^{}_{23}, \delta, \rho, \sigma, m^{}_\beta, m^{}_{\beta\beta}  \}$ at $\Lambda^{}_{\rm EW}$ together with the minimum value $\chi^2_{\rm min}$ of $\chi^2$. The third and fourth columns show results in the Majorana case for the normal and inverted neutrino mass orderings, respectively, whereas the fifth and sixth columns present those in the Dirac case, where $\{\rho,\sigma,m^{}_{\beta\beta}\}$ are not physical observables.}
	\label{tab:best-fits}
\end{table}

We list in Table~\ref{tab:best-fits} the initial inputs for $\{ m^{}_{\rm lightest}, \Delta m^2_{21}, \Delta m^2_{3\ell}, \theta, \phi  \}$ at the high energy scale $\Lambda$ that minimize $\chi^2$, together with the predictions for $\{ m^{}_{\rm lightest}, \Delta m^2_{21}, \Delta m^2_{3\ell}, \theta^{}_{12}, \theta^{}_{13},\theta^{}_{23}, \delta, \rho, \sigma, m^{}_\beta, m^{}_{\beta\beta}  \}$ at the electroweak scale $\Lambda^{}_{\rm EW}$, and the minimum value $\chi^2_{\rm min}$ of $\chi^2$. To clearly illustrate the allowed parameter space and correlations among observables, we show in Figs.~\ref{fig:nmo} and \ref{fig:imo} several representative pairs of predicted observables at $\Lambda^{}_{\rm EW}$ , namely $\left( \theta^{}_{12}, \theta^{}_{13} \right) $, $\left( \delta, \theta^{}_{23} \right) $, $\left( \sigma, \rho \right) $, $\left( \theta^{}_{12}, m^{}_{\rm lightest} \right) $, $\left( m^{}_\beta, m^{}_{\rm lightest}  \right) $, and $\left( m^{}_{\beta\beta}, m^{}_{\rm lightest} \right) $, for the NMO and IMO cases, respectively. In each plot, the cyan star, and the red and blue regions correspond to $\chi^2_{\rm min}$, $\Delta \chi^2 \equiv \chi^2 - \chi^2_{\rm min}\leq 2.30$, and $\Delta \chi^2 \leq 11.83$ for two degrees of freedom, respectively. The lighter grey line in the $\left( m^{}_\beta, m^{}_{\rm lightest}  \right) $ plane and lighter grey band in the $\left( m^{}_{\beta\beta}, m^{}_{\rm lightest}  \right) $ plane denote the upper limits on $m^{}_\beta$ and $m^{}_{\beta\beta}$ from KATRIN and KamLAND-Zen, respectively. The darker grey regions in these two planes represent the allowed $3\sigma$ regions obtained by varying neutrino oscillation parameters within their $3\sigma$ ranges from NuFIT~6.1 and taking $\rho,\sigma \in \left[ 0, 180^\circ \right) $. The dashed line in the $\left( \theta^{}_{12}, \theta^{}_{13} \right)$ plane corresponds to $f(\Lambda) =0$, i.e., the TM1 $\theta^{}_{12}$-$\theta^{}_{13}$ correlation in Eq.~\eqref{eq:TM-relations}, and the one in the $\left( \theta^{}_{12}, m^{}_{\rm lightest} \right)$ plane indicates the maximal value of $\theta^{}_{12}$ predicted by the TM1 $\theta^{}_{12}$-$\theta^{}_{13}$ correlation when $\theta^{}_{12}$ and $\theta^{}_{13}$ vary within their $3\sigma$ region from the NuFIT 6.1 results. We also draw several dashed grey grid-lines (namely, $\delta = 90^\circ, 180^\circ, 270^\circ$ and $\theta^{}_{23} = 45^\circ$) in the $\left( \delta, \theta^{}_{23} \right) $ plane to help visualize the correlation between the quadrant of $\delta$ and the octant of  $\theta^{}_{23}$. Some comments and discussions on these numerical results are given as blow:
\begin{figure}[t!]
	\centering
	\begin{subfigure}{0.32\textwidth}
		\centering
		\includegraphics[width=\linewidth]{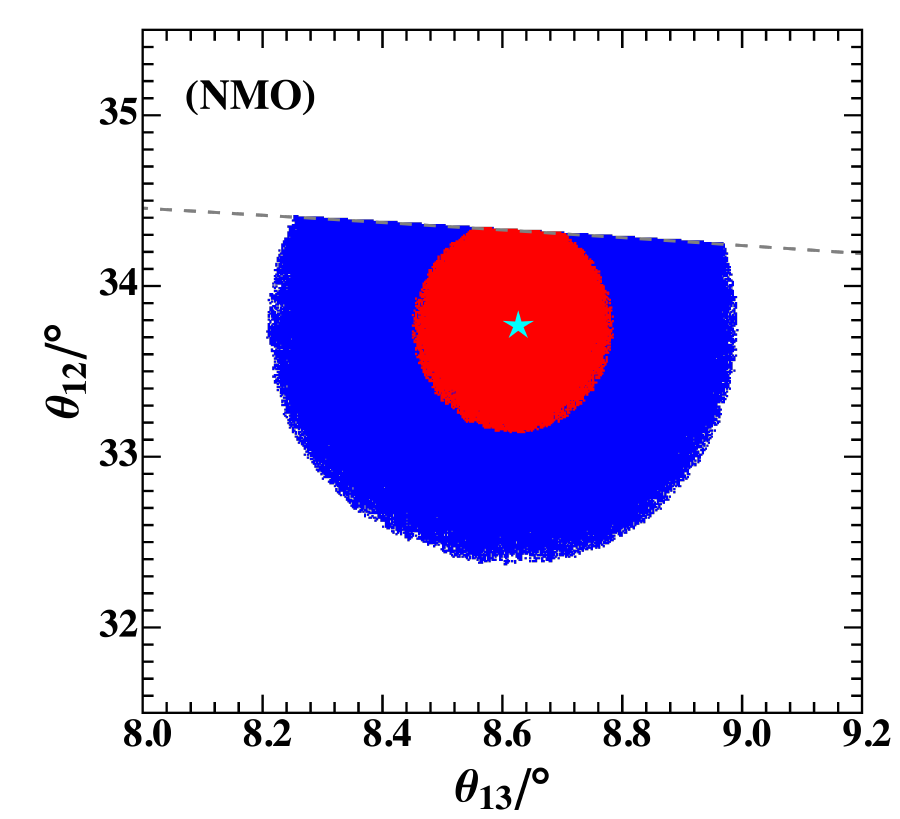}
		%\caption{}
		%\label{fig:sub-a-grid}
	\end{subfigure}
	\hfill
	\begin{subfigure}{0.32\textwidth}
		\centering
		\includegraphics[width=\linewidth]{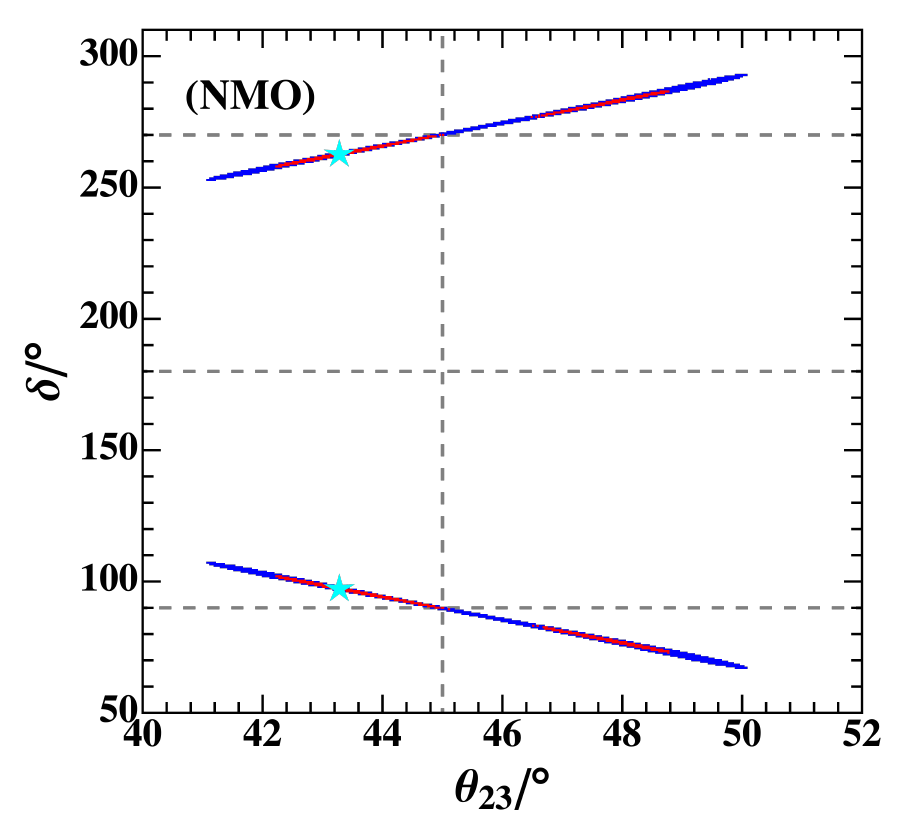}
		%\caption{}
		%\label{fig:sub-b-grid}
	\end{subfigure}
	\hfill
	\begin{subfigure}{0.32\textwidth}
		\centering
		\includegraphics[width=\linewidth]{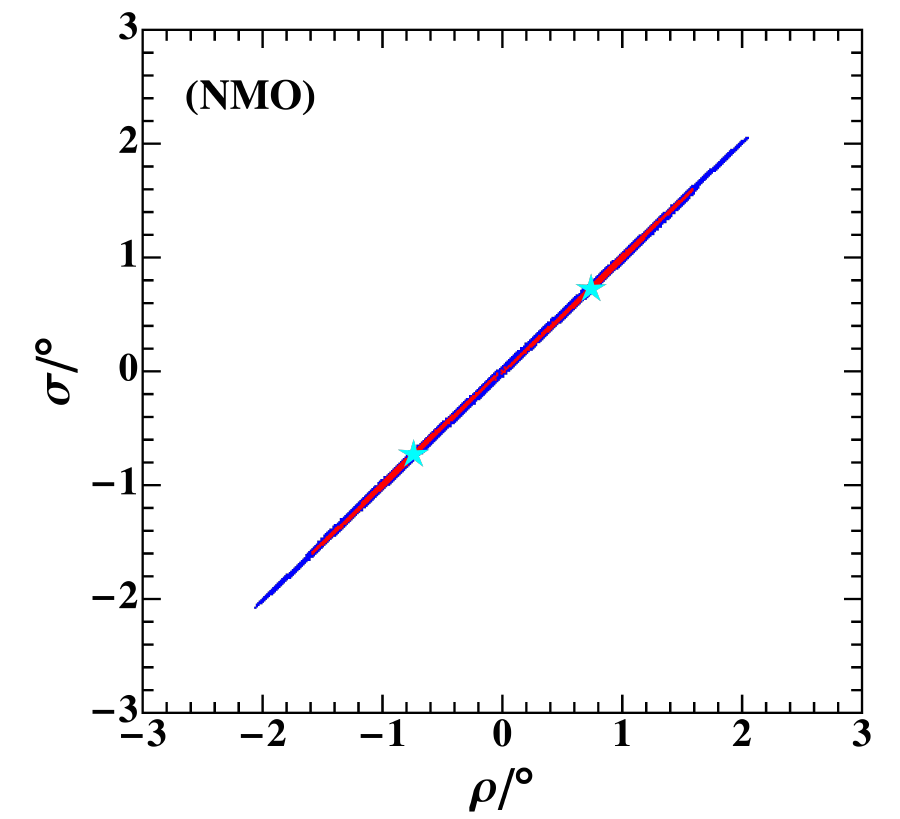}
		%\caption{}
		%\label{fig:sub-c-grid}
	\end{subfigure}
	\\
	\begin{subfigure}{0.32\textwidth}
		\centering
		\includegraphics[width=\linewidth]{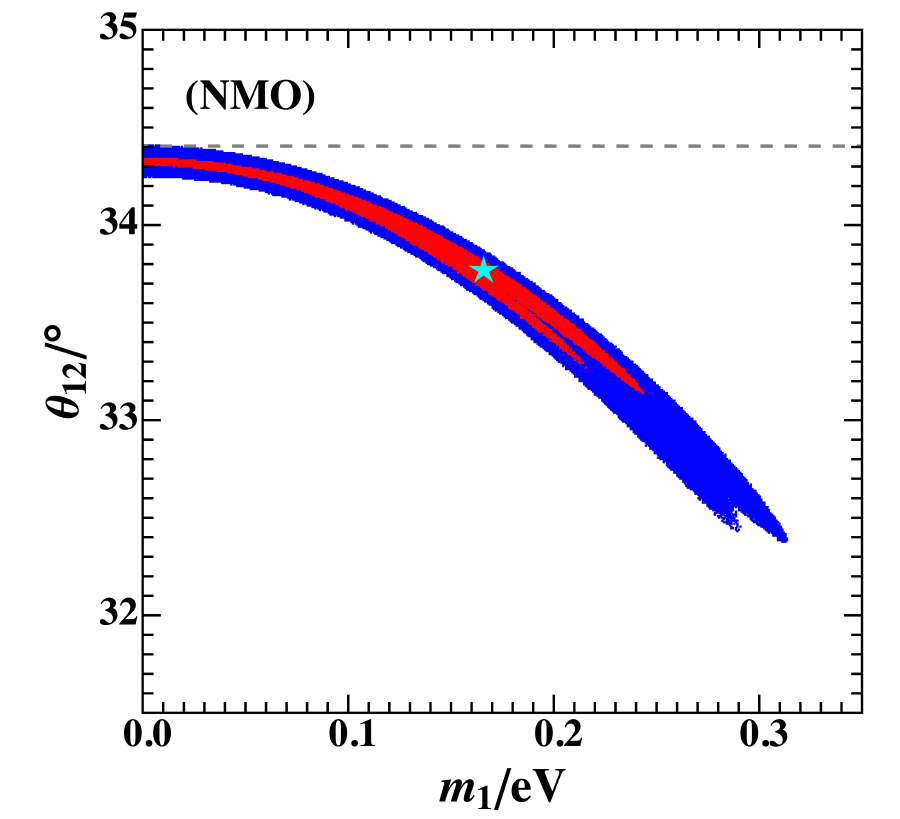} 
		%\caption{}
		%\label{fig:sub-d-grid}
	\end{subfigure}
	\hfill
	\begin{subfigure}{0.32\textwidth}
		\centering
		\includegraphics[width=\linewidth]{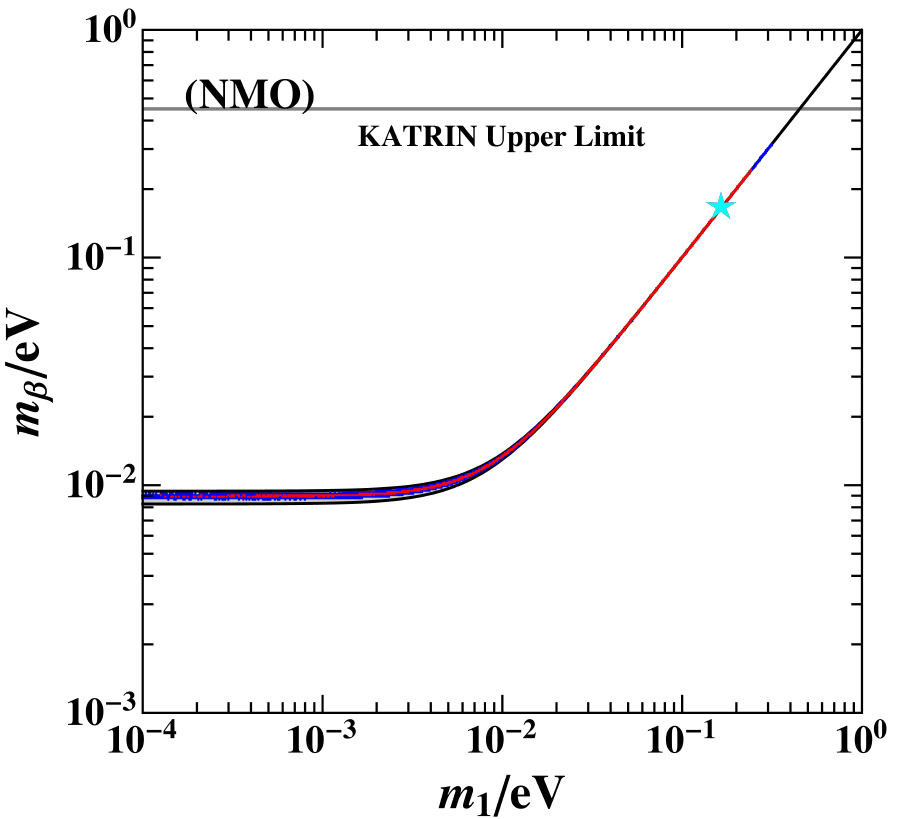}
		%\caption{}
		%\label{fig:sub-e-grid}
	\end{subfigure}
	\hfill
	\begin{subfigure}{0.32\textwidth}
		\centering
		\includegraphics[width=\linewidth]{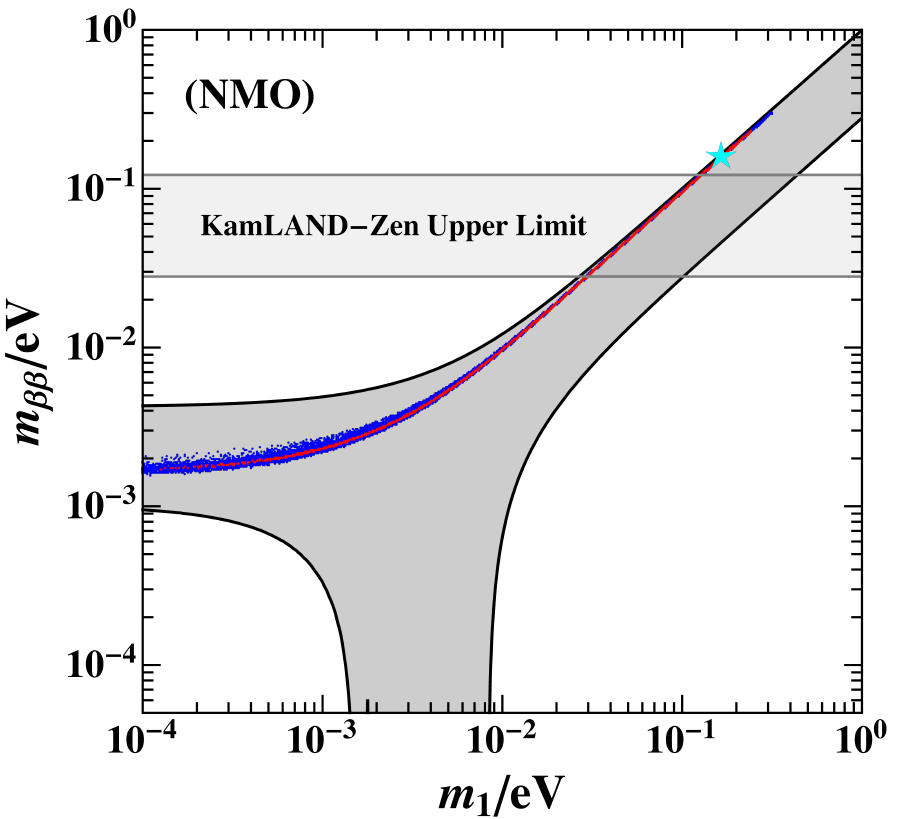}
		%\caption{ }
		%\label{fig:sub-f-grid}
	\end{subfigure}
	\caption{Correlations among observables at $\Lambda^{}_{\rm EW}$ in the NMO case with Majorana neutrinos. The cyan star and the red and blue regions correspond to $\chi^2 = \chi^2_{\rm min}$, $\Delta \chi^2 \leq 2.30$, and $\Delta \chi^2 \leq 11.83$, respectively. Further details are provided in the main text.}
	\label{fig:nmo}
\end{figure}
\begin{figure}[h!]
	\centering
	\begin{subfigure}{0.32\textwidth}
		\centering
		\includegraphics[width=\linewidth]{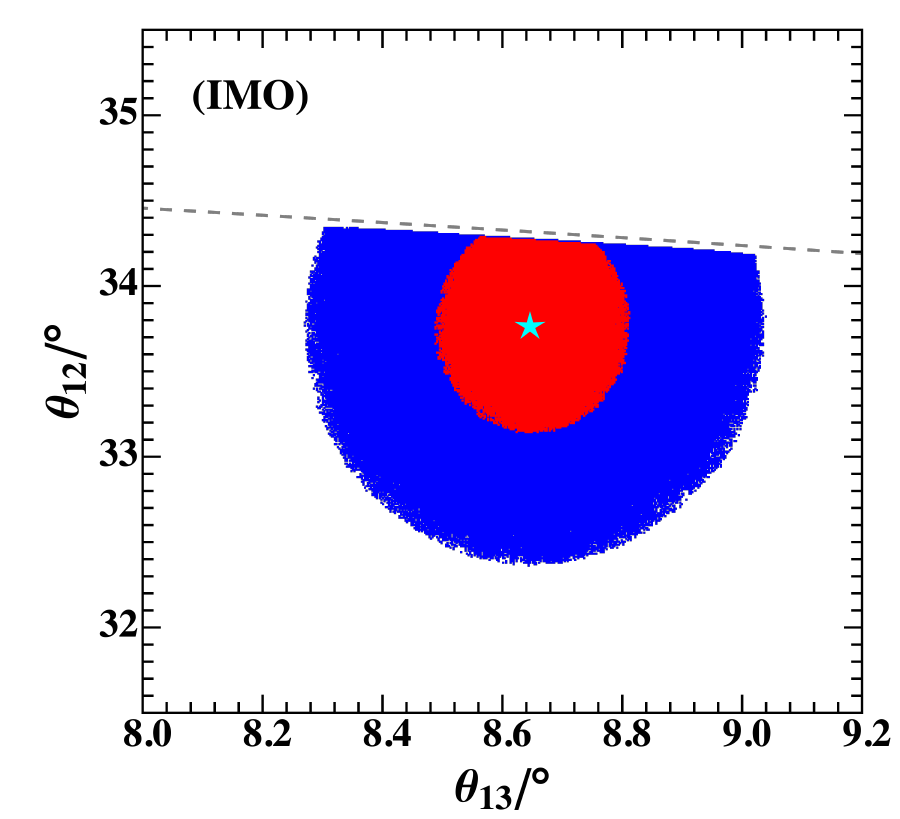}
		%\caption{}
		%\label{fig:sub-a-grid}
	\end{subfigure}
	\hfill
	\begin{subfigure}{0.32\textwidth}
		\centering
		\includegraphics[width=\linewidth]{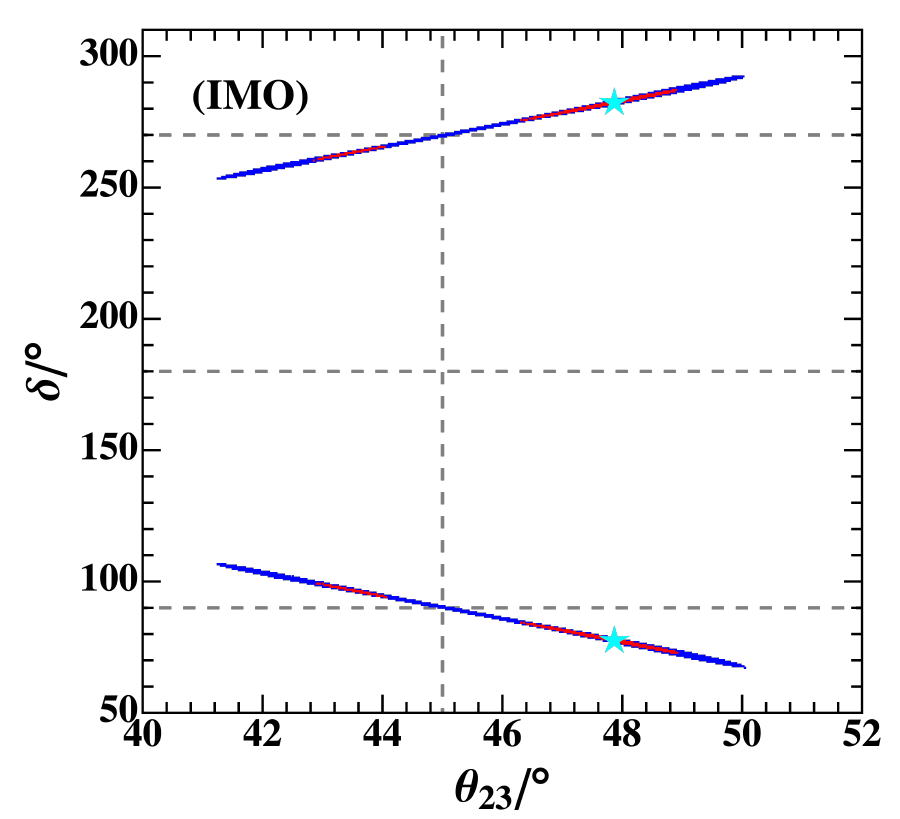}
		%\caption{}
		%\label{fig:sub-b-grid}
	\end{subfigure}
	\hfill
	\begin{subfigure}{0.32\textwidth}
		\centering
		\includegraphics[width=\linewidth]{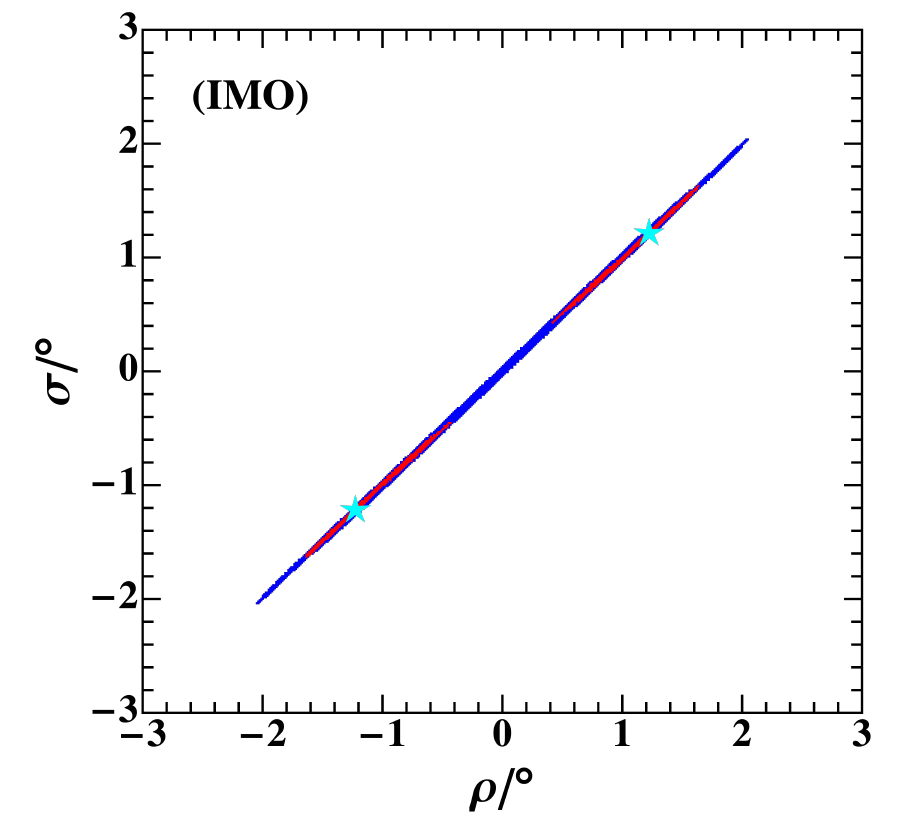}
		%\caption{}
		%\label{fig:sub-c-grid}
	\end{subfigure}
	\\
	\begin{subfigure}{0.32\textwidth}
		\centering
		\includegraphics[width=\linewidth]{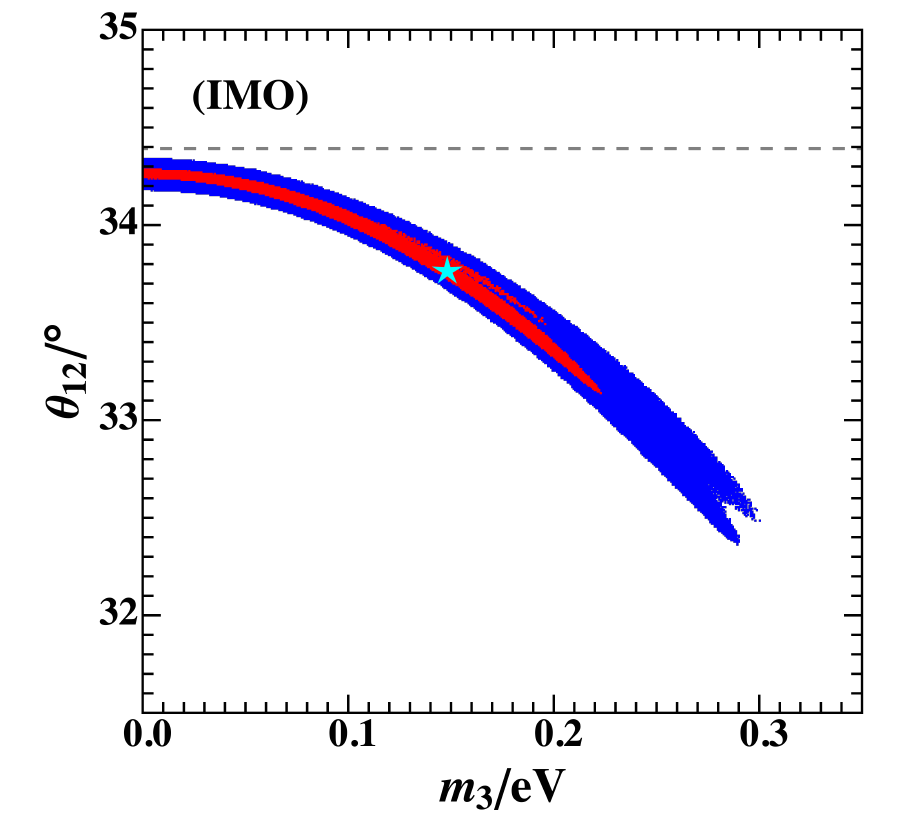} 
		%\caption{}
		%\label{fig:sub-d-grid}
	\end{subfigure}
	\hfill
	\begin{subfigure}{0.32\textwidth}
		\centering
		\includegraphics[width=\linewidth]{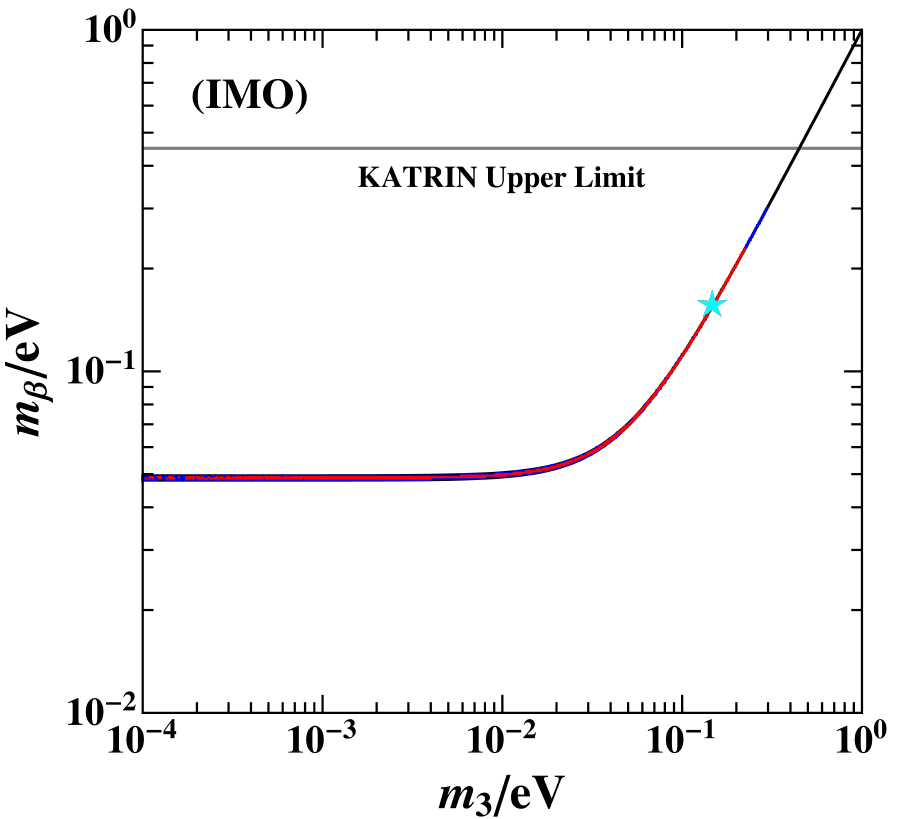}
		%\caption{}
		%\label{fig:sub-e-grid}
	\end{subfigure}
	\hfill
	\begin{subfigure}{0.32\textwidth}
		\centering
		\includegraphics[width=\linewidth]{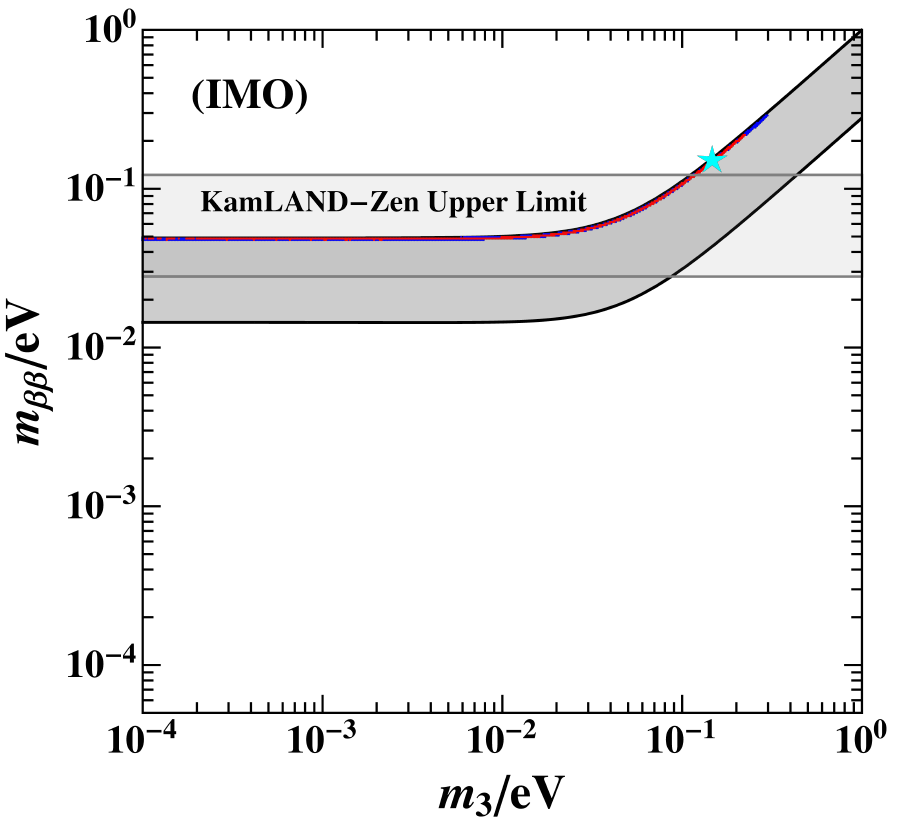}
		%\caption{ }
		%\label{fig:sub-f-grid}
	\end{subfigure}
	\caption{The same as Fig.~\ref{fig:nmo} but in the IMO case.}
	\label{fig:imo}
\end{figure}

\begin{itemize}
	\item As indicated in Table~\ref{tab:best-fits}, the RG corrections can render the TM1 mixing pattern perfectly consistent with the current neutrino oscillation data, with the minimum value of $\chi^2$ being around $\mathcal{O} \left(10^{-2} \right)$ in both the NMO and IMO cases. However, achieving such agreement requires a nearly degenerate mass spectrum with $m^{}_{\rm lightest} \sim 0.15$~eV. This is consistent with the analytical discussions in the previous section, and moreover, the value of $m^{}_1$ has already been well estimated with the help of Eq.~\eqref{eq:fun-app}.
	
	\item The upper-left panels in Figs.~\ref{fig:nmo} and \ref{fig:imo} show that the RG corrections to $\theta^{}_{12}$ fortunately decrease its value and are able to make $\theta^{}_{12}$ and $\theta^{}_{13}$ distributed within their allowed $1\sigma$ and $3\sigma$ regions below the $f\left( \Lambda \right) =0$ line in both the NMO and IMO cases. Note that a visible gap exists between the $f\left( \Lambda \right) =0$ line and the $3\sigma$ region in the IMO case, which does not appear in the NMO case. This feature is attributed to the strong degeneracy between $m^{}_{1}$ and $m^{}_2$ even when $m^{}_3 = 0$ in the IMO case, and hence $f\left( \Lambda^{}_{\rm EW} \right)$ is always enhanced by $\zeta^{-1}_{21}$ and deviates from $f\left( \Lambda \right)=0$. In contrast, in the NMO case with $m^{}_1 = 0$, no such enhancement occurs since $\zeta^{-1}_{21} = 1$ and the contribution from $\zeta^{-1}_{32} $ is suppressed by $\sin\theta^{}_{13}$, as indicated in Eq.~\eqref{eq:rge-th12-M}. This behavior is also reflected in the lower-left panels of Figs.~\ref{fig:nmo} and \ref{fig:imo}, which further indicate that a larger $m^{}_1$ ($m^{}_3$)  leads to a smaller $\theta^{}_{12}$. As a result, avoiding a too small value of $\theta^{}_{12}$ imposes an upper limit on the lightest neutrino mass, i.e., $m^{}_1 \lesssim 0.313$~eV  ($m^{}_3 \lesssim 0.304$~eV) in the NMO (IMO) case.
	
	\item The upper-middle panels in Figs.~\ref{fig:nmo} and \ref{fig:imo} display two separated regions for the Dirac CP-violating phase $\delta$, whose best-fit values are $\delta = 97.295^\circ$ and $ 262.70^\circ$ in the NMO case, and $\delta = 77.605^\circ$ and $ 282.40^\circ$ in the IMO case. These two separated regions arise from the fact that both $\delta$ and $2\pi - \delta$ can be extracted from $\cos\delta$, and only $\cos\delta$ is determined by $\theta^{}_{23}$ and $\theta^{}_{13}$ in the TM1 mixing pattern, which can be seen from the relation in Eq.~\eqref{eq:TM1-relations}. Furthermore, it shows that the quadrant of $\delta$ is still tightly correlated with the octant of $\theta^{}_{23}$: the first octant of $\theta^{}_{23}$ corresponds to the second and third quadrants of $\delta$, while the second octant of $\theta^{}_{23}$ corresponds to the first and fourth quadrants of $\delta$. As shown in the upper-right panels of Figs.~\ref{fig:nmo} and \ref{fig:imo}, the two Majorana phases $\rho$ and $\sigma$ also appear in two separated regions around zero. This behavior arises because they are roughly determined by $\rho=\sigma = \phi - \delta$, where the difference between $\phi$ and $\delta$ approximately governed by $\sin\phi = \sin2\theta^{}_{23} \sin\delta$ is small.
				
	\item As discussed above, a large lightest neutrino mass is required to generate sizable RG corrections to $\theta^{}_{12}$. However, such a large mass also results in large values of both $m^{}_\beta$ and $m^{}_{\beta\beta}$, which may be strongly constrained by current experimental limits. From the lower-middle and lower-right panels of Figs.~\ref{fig:nmo} and \ref{fig:imo}, one can see that the predicted $3\sigma$ range of $m^{}_\beta$ is well below the latest KATRIN upper bound in both the NMO and IMO cases. In contrast, the predicted best-fit value and part of the $1\sigma$ range of $m^{}_{\beta\beta}$  already exceed the KamLAND-Zen upper bound. Compared with the NMO case where part of the $1\sigma$ range of $m^{}_{\beta\beta}$ still lies below the upper limit, the IMO case faces a stronger challenge from neutrinoless double beta decay experiments, as its remaining $1\sigma$ and $3\sigma$ ranges lie entirely within the experimental uncertainty.
\end{itemize}

\begin{figure}[t!]
	\centering
	\begin{subfigure}{0.32\textwidth}
		\centering
		\labellist
		\tiny
		\pinlabel {with $\rho^\prime, \sigma^\prime$} [l] at 160 348
		\endlabellist
		\includegraphics[width=\linewidth]{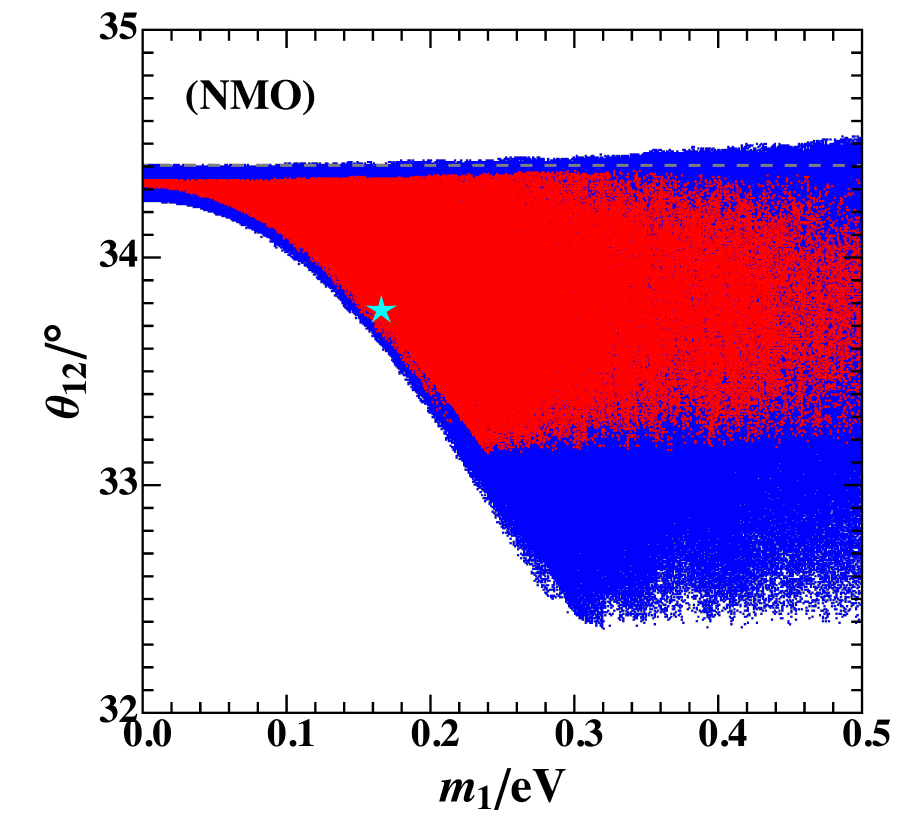} 
		%		\caption{}
		%		\label{fig:sub-d-grid}
	\end{subfigure}
	\hfill
	\begin{subfigure}{0.32\textwidth}
		\centering
		\labellist
		\tiny
		\pinlabel {with $\rho^\prime$} [l] at 160 348
		\endlabellist
		\includegraphics[width=\linewidth]{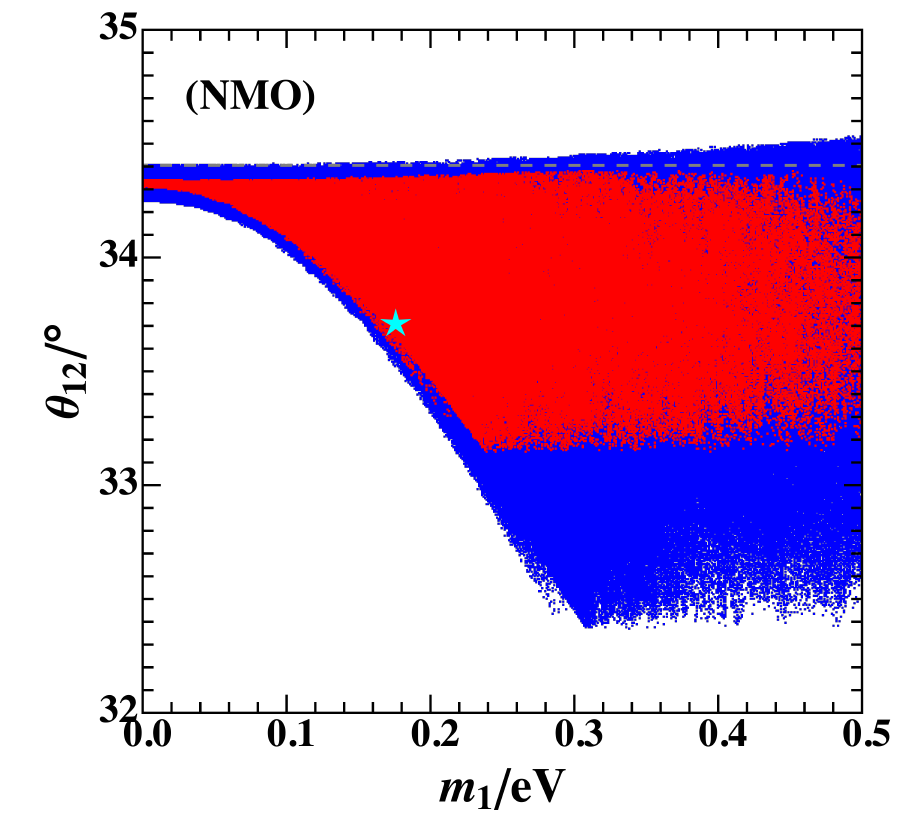}
		%		\caption{}
		%		\label{fig:sub-e-grid}
	\end{subfigure}
	\hfill
	\begin{subfigure}{0.32\textwidth}
		\centering
		\labellist
		\tiny
		\pinlabel {with $\sigma^\prime$} [l] at 160 348
		\endlabellist
		\includegraphics[width=\linewidth]{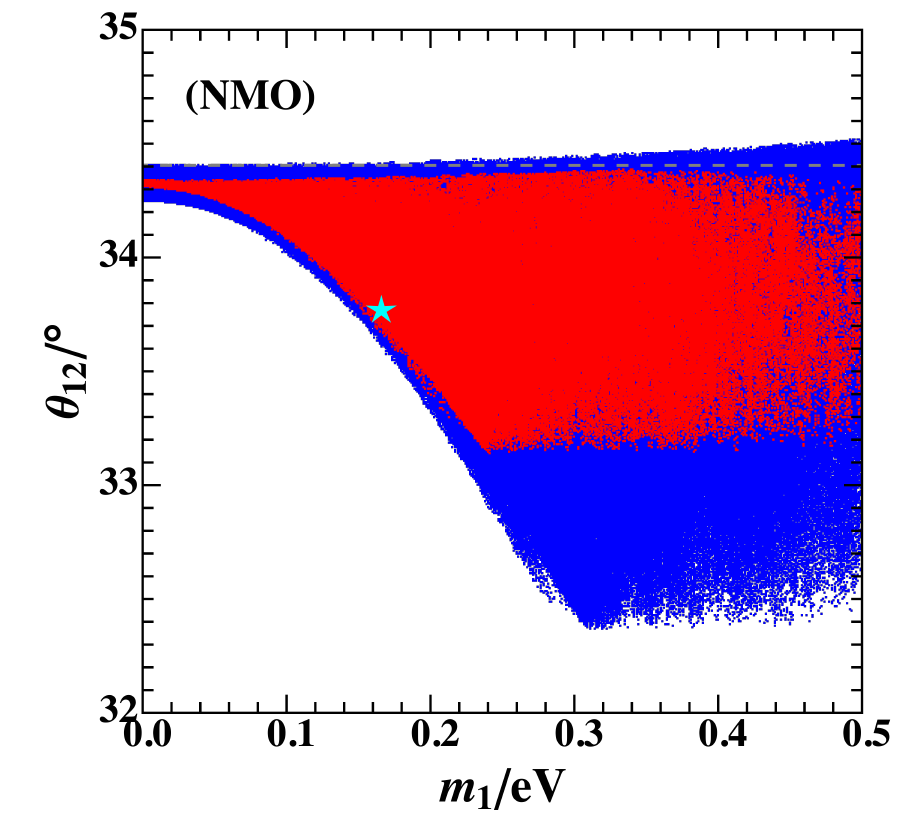}
		%		\caption{ }
		%		\label{fig:sub-f-grid}
	\end{subfigure}
	\\
	\begin{subfigure}{0.32\textwidth}
		\centering
		\labellist
		\tiny
		\pinlabel {with $\rho^\prime, \sigma^\prime$} [l] at 160 348
		\endlabellist
		\includegraphics[width=\linewidth]{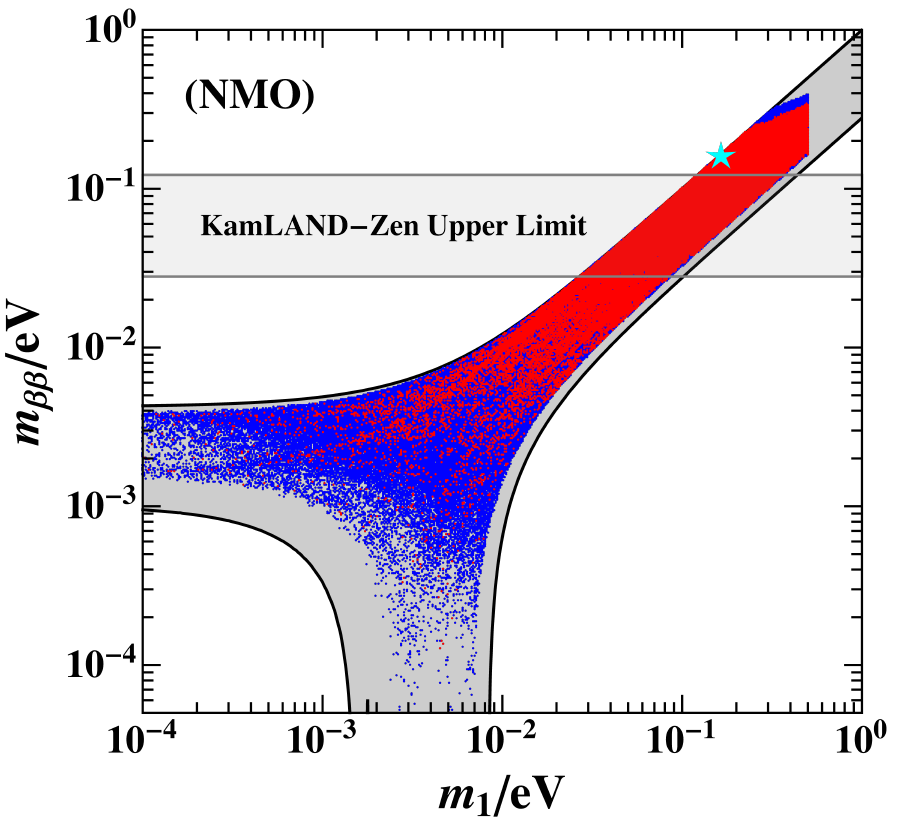} 
		%		\caption{}
		%		\label{fig:sub-d-grid}
	\end{subfigure}
	\hfill
	\begin{subfigure}{0.32\textwidth}
		\centering
		\labellist
		\tiny
		\pinlabel {with $\rho^\prime$} [l] at 160 348
		\endlabellist
		\includegraphics[width=\linewidth]{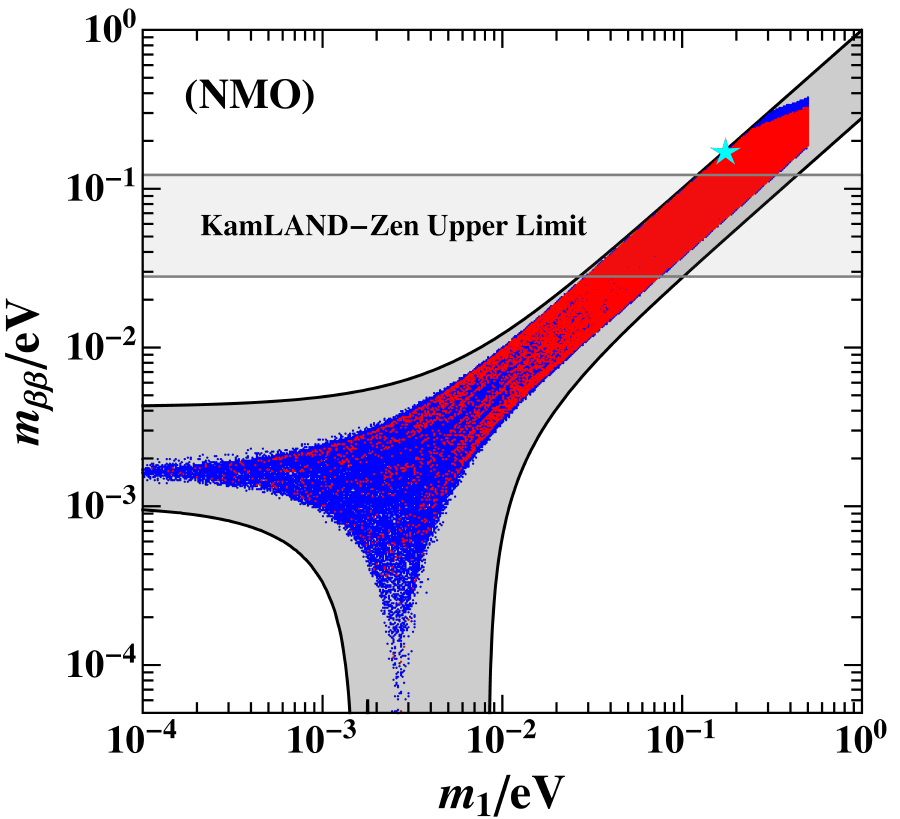}
		%		\caption{}
		%		\label{fig:sub-e-grid}
	\end{subfigure}
	\hfill
	\begin{subfigure}{0.32\textwidth}
		\centering
		\labellist
		\tiny
		\pinlabel {with $\sigma^\prime$} [l] at 160 348
		\endlabellist
		\includegraphics[width=\linewidth]{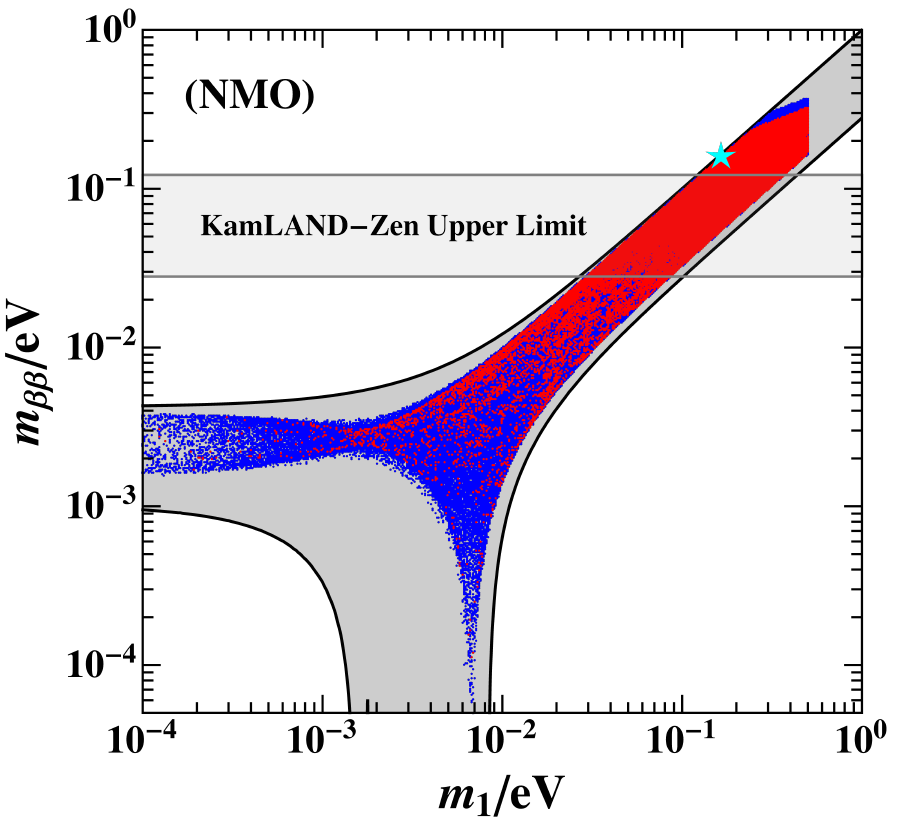}
		%		\caption{ }
		%		\label{fig:sub-f-grid}
	\end{subfigure}
	\caption{The same as Fig.~\ref{fig:nmo} but with $\rho^\prime$ and $\sigma^\prime$ or with only one of them in the NMO case.}
	\label{fig:mps}
\end{figure}

\begin{figure}[t!]
	\centering
	\begin{subfigure}{0.32\textwidth}
		\centering
		\labellist
		\tiny
		\pinlabel {with $\rho^\prime, \sigma^\prime$} [l] at 160 348
		\endlabellist
		\includegraphics[width=\linewidth]{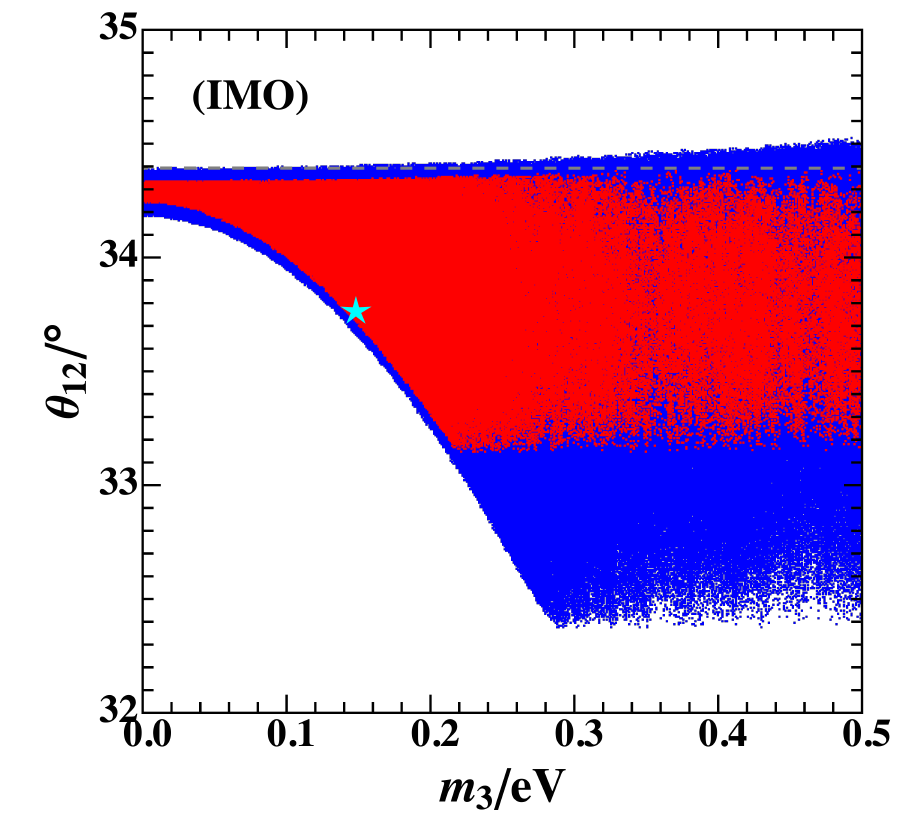} 
		%		\caption{}
		%		\label{fig:sub-d-grid}
	\end{subfigure}
	\hfill
	\begin{subfigure}{0.32\textwidth}
		\centering
		\labellist
		\tiny
		\pinlabel {with $\rho^\prime$} [l] at 160 348
		\endlabellist
		\includegraphics[width=\linewidth]{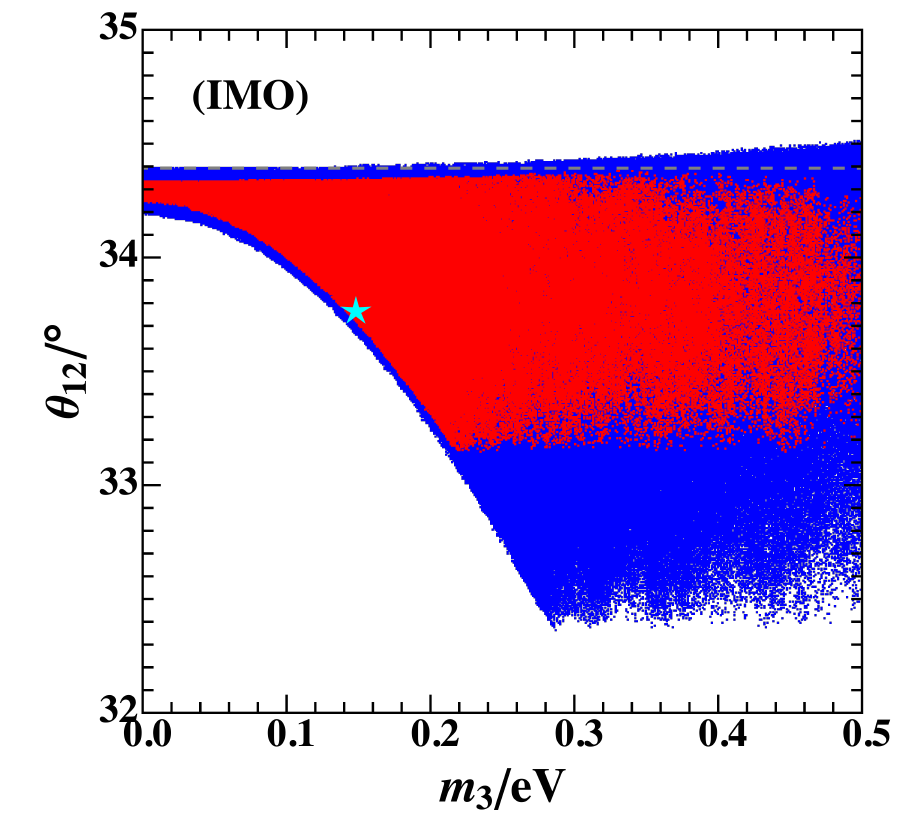}
		%		\caption{}
		%		\label{fig:sub-e-grid}
	\end{subfigure}
	\hfill
	\begin{subfigure}{0.32\textwidth}
		\centering
		\labellist
		\tiny
		\pinlabel {with $\sigma^\prime$} [l] at 160 348
		\endlabellist
		\includegraphics[width=\linewidth]{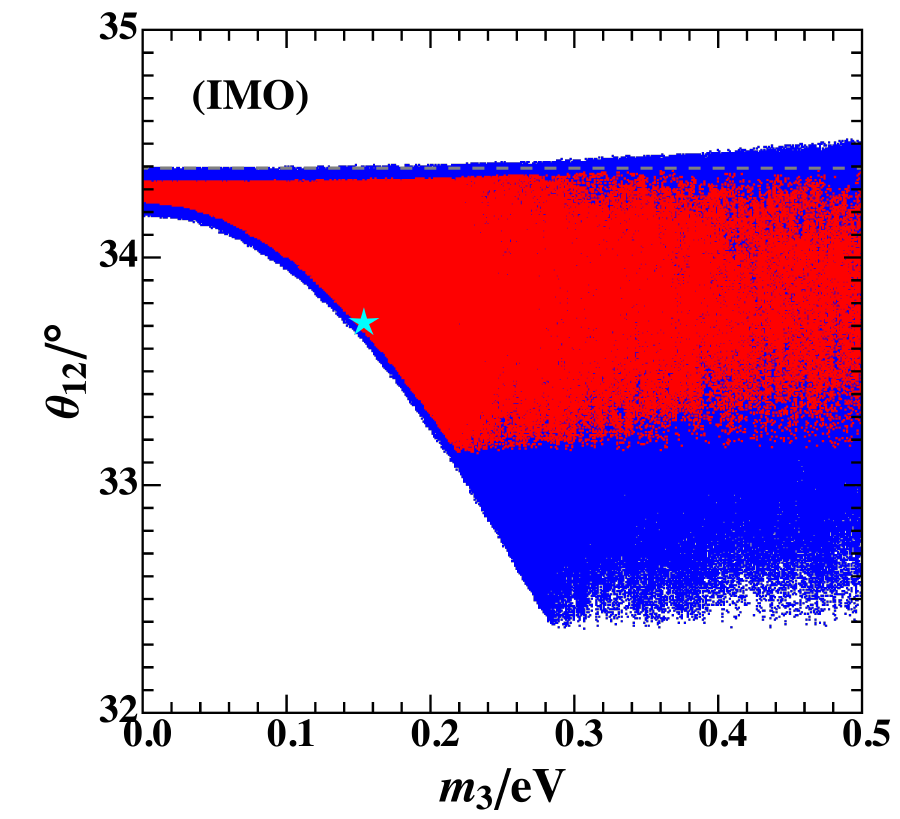}
		%		\caption{ }
		%		\label{fig:sub-f-grid}
	\end{subfigure}
	\\
	\begin{subfigure}{0.32\textwidth}
		\centering
		\labellist
		\tiny
		\pinlabel {with $\rho^\prime, \sigma^\prime$} [l] at 160 348
		\endlabellist
		\includegraphics[width=\linewidth]{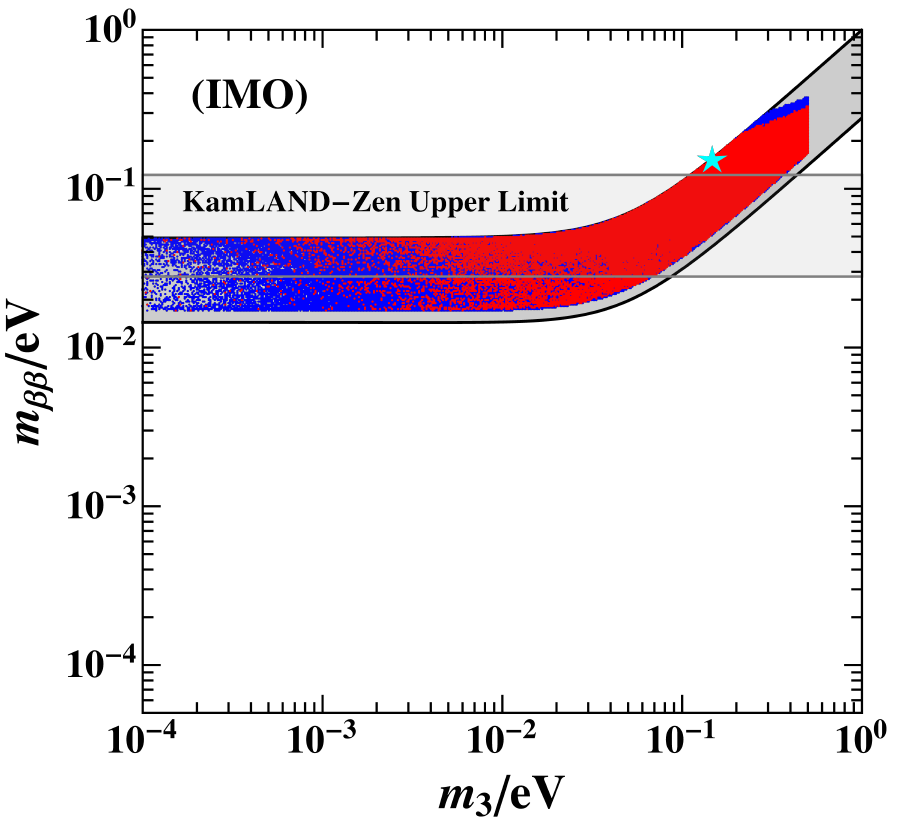} 
		%		\caption{}
		%		\label{fig:sub-d-grid}
	\end{subfigure}
	\hfill
	\begin{subfigure}{0.32\textwidth}
		\centering
		\labellist
		\tiny
		\pinlabel {with $\rho^\prime$} [l] at 160 348
		\endlabellist
		\includegraphics[width=\linewidth]{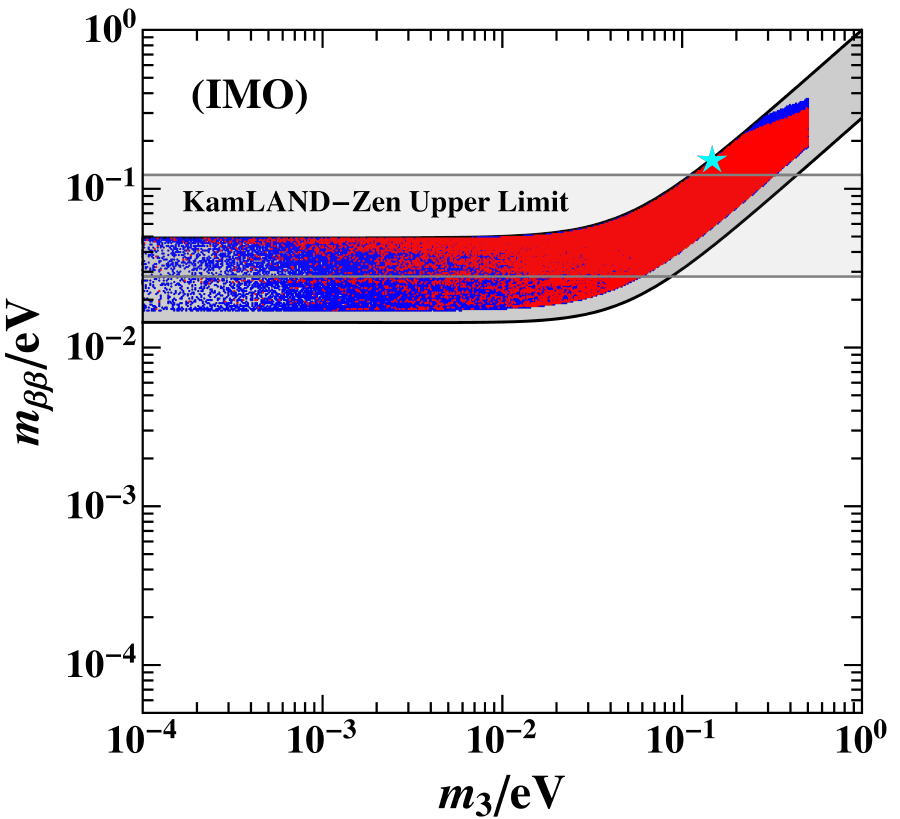}
		%		\caption{}
		%		\label{fig:sub-e-grid}
	\end{subfigure}
	\hfill
	\begin{subfigure}{0.32\textwidth}
		\centering
		\labellist
		\tiny
		\pinlabel {with $\sigma^\prime$} [l] at 160 348
		\endlabellist
		\includegraphics[width=\linewidth]{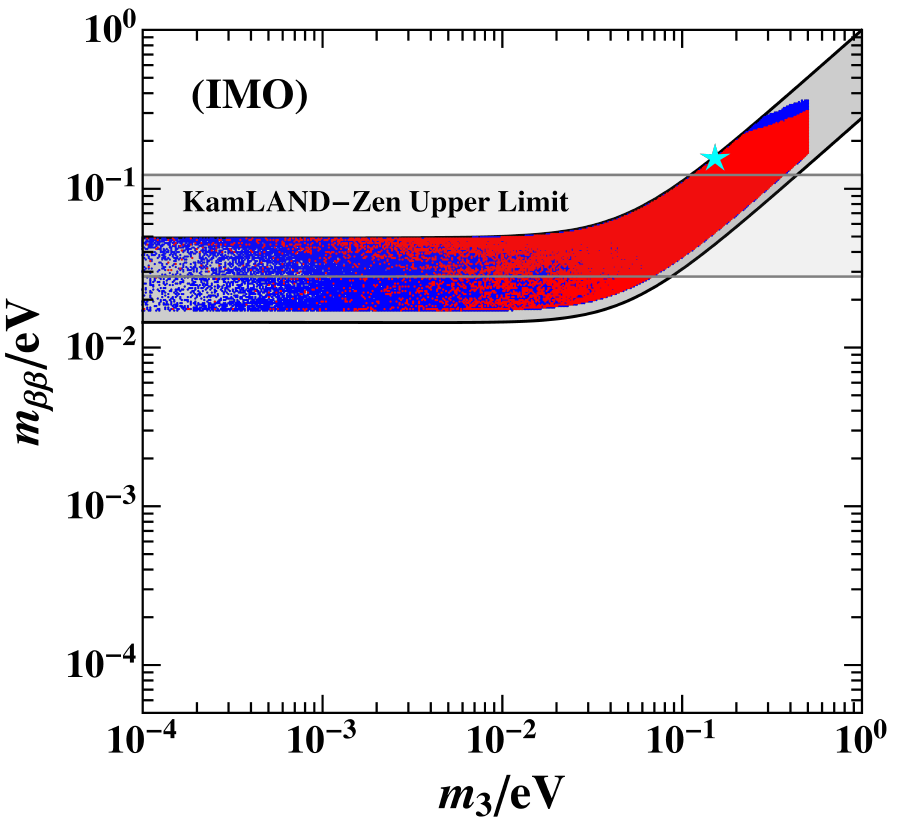}
		%		\caption{ }
		%		\label{fig:sub-f-grid}
	\end{subfigure}
	\caption{The same as Fig.~\ref{fig:nmo} but with $\rho^\prime$ and $\sigma^\prime$ or with only one of them in the IMO case.}
	\label{fig:mps-imo}
\end{figure}

It is worth noting that all the above analyses are based on the pure TM1 mixing matrix $U^{}_{\rm TM1}$ given in Eq.~\eqref{eq:TM1}. In general, an additional diagonal phase matrix $P={\rm Diag} \left\{ e^{\rmi \rho^\prime}, e^{\rmi \sigma^\prime}, 1\right\}$, rephasing Majorana neutrino fields, can be introduced on the right-hand side of $U^{}_{\rm TM1}$. The resulting mixing matrix $U^\prime_{\rm TM1} = U^{}_{\rm TM1} P $ yields the same relations as those derived from $U^{}_{\rm TM1}$ for all mixing parameters except the two Majorana phases, before including the RG corrections. With this new mixing matrix $U^\prime_{\rm TM1}$, the two Majorana phases become
\begin{eqnarray}
	\rho = \phi - \delta + \rho^\prime \;,\quad \sigma = \phi - \delta + \sigma^\prime \;,
\end{eqnarray}
and now they behave as free parameters and are allowed to vary over the entire range $\left[ 0, 180^\circ \right)$. Consequently, the predicted region of $m^{}_{\beta\beta}$ can be significantly modified, and the RG corrections may also be strongly affected by the widely varying Majorana phases. However, the key question is whether these effects can alleviate the constraint from the neutrinoless double beta decay experiments. To examine this point, we perform the same numerical calculations as before but include two additional input parameters, $\rho^\prime$ and $\sigma^\prime$, besides $\{ m^{}_{\rm lightest}, \Delta m^2_{21}, \Delta m^2_{3\ell}, \theta, \phi  \}$. It is unnecessary to minimize $\chi^2$ again, and instead, we directly take the previous best-fit initial inputs for $\{ m^{}_{\rm lightest}, \Delta m^2_{21}, \Delta m^2_{3\ell}, \theta, \phi  \}$ given in Table~\ref{tab:best-fits}, together with $\rho^\prime = \sigma^\prime = 0$, as the starting point for the Metropolis algorithm to explore the allowed parameter space. During the search process, we further restrict $m^{}_{\rm lightest} \left( \Lambda^{}_{\rm EW} \right) \lesssim 0.5$ eV to avoid generating excessively large $m^{}_{\rm lightest}$ and a consequent reduction in the sampling efficiency. The results for correlations of $\theta^{}_{12}$ and $m^{}_{\beta\beta}$ with $m^{}_{\rm lightest}$ are presented in Figs.~\ref{fig:mps} and \ref{fig:mps-imo} for three cases where both $\rho^\prime$ and $\sigma^\prime$, or only one of them is introduced. As expected, the allowed $1\sigma$ and $3\sigma$ regions for the pair of $m^{}_{\beta\beta}$ and $m^{}_{\rm lightest}$ are enlarged a lot. In particular, part of $1\sigma$ and $3\sigma$ regions now lie below the KamLAND-Zen upper limit in both the NMO and IMO cases, although the best-fit points remain above the  limit. Note that, in specific flavor models leading to the TM mixing patterns~\cite{Grimus:2008tt,deMedeirosVarzielas:2012apl,Luhn:2013lkn,Li:2013jya,Zhao:2014yaa,Girardi:2015rwa,Petcov:2018snn,Novichkov:2018yse,King:2019vhv,Ding:2020vud,Thapa:2021ehj,deMedeirosVarzielas:2021pug,Zhang:2024rwv}, one or two Majorana phases may be correlated with other model parameters and hence are not necessarily free. The number of independent phases depends on the detailed structure of the underlying flavor symmetry and its breaking pattern. Consequently, the model-independent results shown in Figs.~\ref{fig:mps} and~\ref{fig:mps-imo} generally overestimate the available parameter space of concrete flavor realizations of the TM1 mixing pattern. Nevertheless, these results provide a useful overview of the qualitative impact of varying Majorana phases.

The upper panels of Fig.~\ref{fig:mps} and \ref{fig:mps-imo} indicate a notable broadening of the allowed $1\sigma$ and $3\sigma$ regions for the $\left( \theta^{}_{12}, m^{}_{\rm lightest} \right)$ pair, which now extend over the entire range of large $m^{}_{\rm lightest}$ values. Furthermore, part of the $3\sigma$ region of $\left( \theta^{}_{12}, m^{}_{\rm lightest} \right)$ even lies above the maximal $\theta^{}_{12}$ value (the dashed grey line) predicted by the TM1 $\theta^{}_{12}$-$\theta^{}_{13}$ correlation within the experimental $3\sigma$ ranges of $\theta^{}_{12}$ and $\theta^{}_{13}$, This means that the direction of the RG corrections to $\theta^{}_{12}$ in this region is opposite to that in the others. These behaviors can be well understood with the help of Eq.~\eqref{eq:rge-th12-M}, which shows that the enhancement factor $\zeta^{-1}_{21}$ is multiplied by $\cos^2\left( \rho - \sigma \right)$ or $\sin2\left( \rho -\sigma \right)$. In the case without $\rho^\prime$ and $\sigma^\prime$, $\rho \simeq \sigma$ holds and leads to $\cos^2\left( \rho - \sigma \right) \simeq 1$ and $\sin2\left( \rho -\sigma \right)=0$, so that the enhancement of $\Delta \theta^{}_{12}$ from large $\zeta^{-1}_{21}$ is unsuppressed.  In contrast, when $\rho^\prime$ and (or) $\sigma^\prime$ are introduced, $\rho$ and (or) $\sigma$ become independent free parameters, and the enhancement from $\zeta^{-1}_{21}$ can be largely suppressed by the small or even vanishing  values of $\cos^2\left( \rho - \sigma \right) $ and $\sin2\left( \rho -\sigma \right)$. When $\rho - \sigma$ approaches $\pm 90^\circ$, a larger $\zeta^{-1}_{21}$ (equivalently a larger $m^{}_{\rm lightest}$) is required to compensate for the suppression by small $\cos^2\left( \rho - \sigma \right) $ and $\sin2\left( \rho -\sigma \right)$. This explains why the allowed $\theta^{}_{12}$-$m^{}_{\rm lightest}$ region can extend across the entire range of large $m^{}_{\rm lightest}$ values. When $\rho - \sigma$ is extremely close or equal to $\pm 90^\circ$, the terms involving $\zeta^{-1}_{12}$ are strongly suppressed or even entirely vanish, and the contributions from $\zeta^{-1}_{31}$ and $\zeta^{-1}_{32}$ become dominant, namely,
\begin{eqnarray}\label{eq:theta12-mps}
	\Delta \theta^{}_{12} \left( \Lambda^{}_{\rm EW} \right) \sim \frac{1}{6} \Delta^{}_\tau \sin\theta^{}_{13} \sin2\theta^{}_{23} \left[ \zeta^{-1}_{31} \cos\delta\cos\left( \delta + 2\rho \right) + 2 \zeta^{-1}_{32} \sin\delta \sin\left( \delta + 2\rho \right) \right] \;,
\end{eqnarray}
where $\rho - \sigma \sim \pm 90^\circ$ and the correlations in Eqs.~\eqref{eq:TM-relations} and \eqref{eq:TM1-relations} have been exploited. It is evident that $\Delta \theta^{}_{12}  \left( \Lambda^{}_{\rm EW} \right) $ can be positive for certain values of $\rho$, e.g., $\rho = 0$. However, the enhancements from $\zeta^{-1}_{31}$ and $\zeta^{-1}_{32}$ are generally much smaller than the unsuppressed contributions from $\zeta^{-1}_{21}$. Therefore, the allowed $\theta^{}_{12}$-$m^{}_{\rm lightest}$ region can lie above the dashed grey line, particularly, in the large $m^{}_{\rm lightest}$ range, but it does not extend far beyond it~\footnote{This RG running behavior of $\theta^{}_{12}$ is reminiscent of RG corrections in the Minimal Supersymmetric Model (MSSM), which can be substantially enhanced by a large $\tan\beta$ and typically proceed in the opposite direction compared with those in the SM~\cite{Zhang:2020lsd,Casas:1999tg,Antusch:2003kp,Antusch:2005gp,Mei:2005qp,Antusch:2001vn}. Due to the opposite RG running direction, RG corrections in the MSSM are not able to bring the TM1 mixing pattern closer to the experimental results, but instead drive it further away, when adopting the pure TM1 mixing matrix $U^{}_{\rm TM1}$. Nevertheless, as indicated by Eq.~\eqref{eq:theta12-mps}, it is possible to reverse the direction of the RG corrections to~$\theta^{}_{12}$ by introducing additional phases $\rho^\prime$ and $\sigma^\prime$. This implies that, within certain regions of parameter space, RG corrections in the MSSM could act in the right direction to reduce $\theta^{}_{12}$. However, to which extent MSSM RG effects with reasonable values of $\tan\beta$ and $m^{}_{\rm lightest}$ can reconcile the TM1 mixing pattern with experimental data remains to be carefully investigated, which is beyond the scope of the present work.}.

In summary, introducing the one or two additional phases from the right-hand side of $U^{}_{\rm TM1}$ can extremely relax the constraint from neutrinoless double beta decay experiments, although the best-fit value and part of the $1\sigma$ range of $m^{}_{\beta\beta}$ still lie above the current experimental upper limit. 

\subsection{The Dirac Case}

In the previous subsection, we have shown that the TM1 mixing pattern in the Majorana case can fully coincide with the latest JUNO result once RG corrections are taken into account, although its parameter space receives strong constraints from neutrinoless double beta decay experiments. Given the fact that RG corrections in the Dirac case exhibit similar enhancement and direction to those in the Majorana case (see, e.g., Eq.~\eqref{eq:fun-app}), it is expected that RG corrections in the Dirac case can not only render the TM1 mixing pattern consistent with the JUNO data but also avoid the strong constraints from neutrinoless double beta decay experiments. 

\begin{figure}[t!]
	\centering
	\begin{subfigure}{0.32\textwidth}
		\centering
		\includegraphics[width=\linewidth]{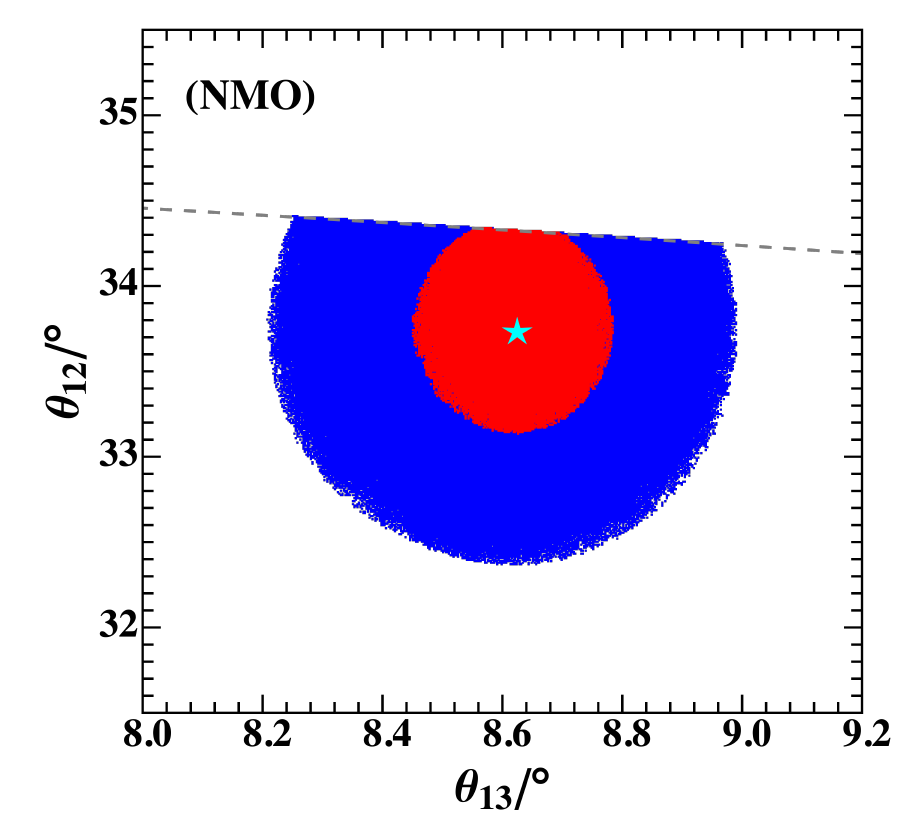}
		%		\caption{}
		%		\label{fig:sub-a-grid}
	\end{subfigure}
	\hfill
	\begin{subfigure}{0.32\textwidth}
		\centering
		\includegraphics[width=\linewidth]{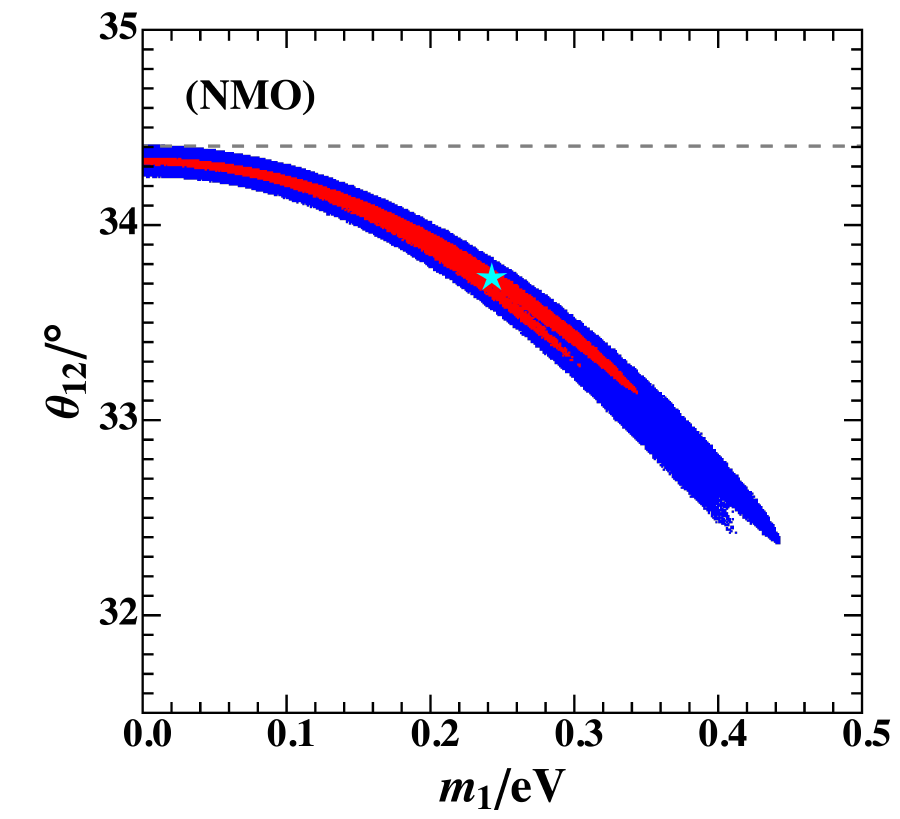} 
		%		\caption{}
		%		\label{fig:sub-d-grid}
	\end{subfigure}
	\hfill
	\begin{subfigure}{0.32\textwidth}
		\centering
		\includegraphics[width=\linewidth]{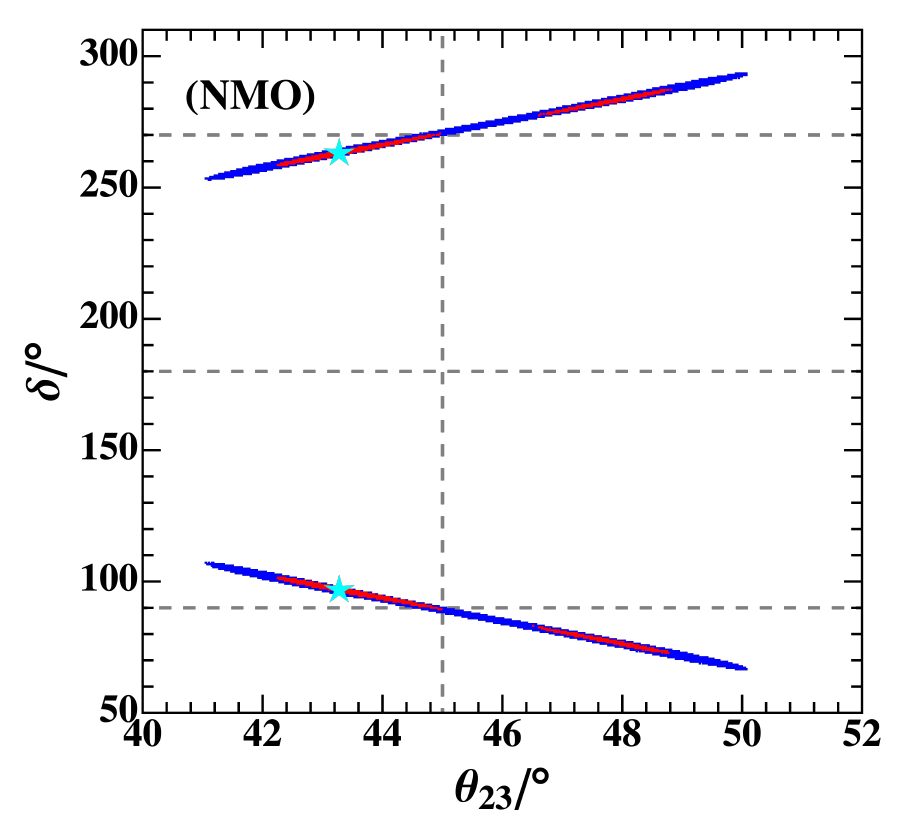}
		%		\caption{}
		%		\label{fig:sub-b-grid}
	\end{subfigure}
	\\
	\begin{subfigure}{0.32\textwidth}
		\centering
		\includegraphics[width=\linewidth]{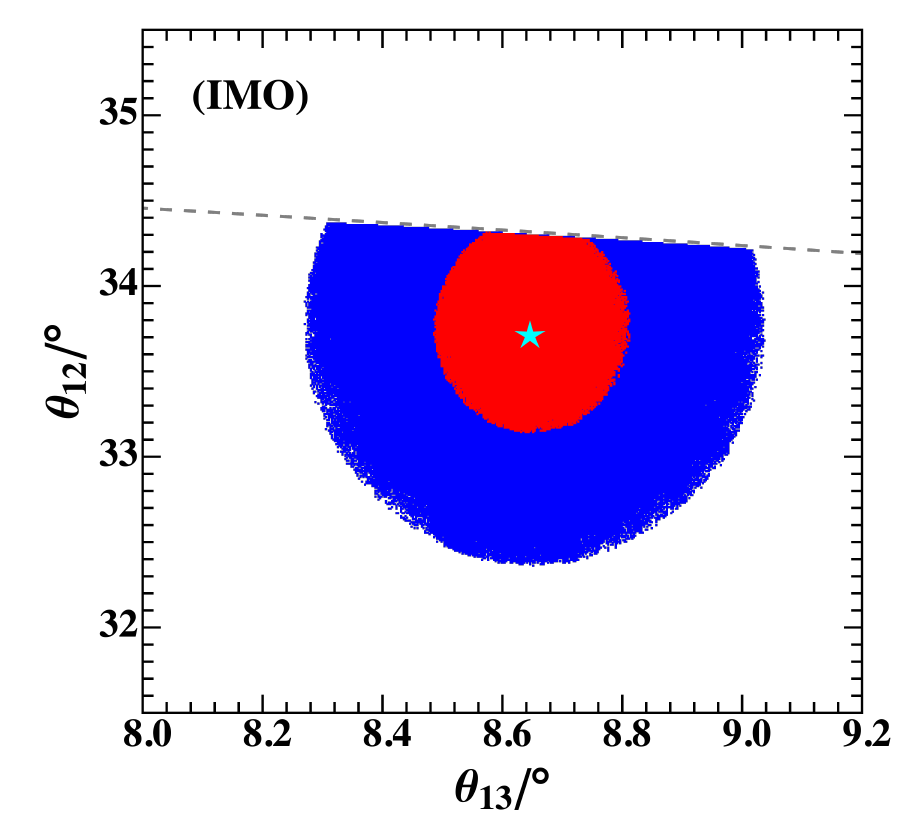}
		%		\caption{}
		%		\label{fig:sub-a-grid}
	\end{subfigure}
	\hfill
	\begin{subfigure}{0.32\textwidth}
		\centering
		\includegraphics[width=\linewidth]{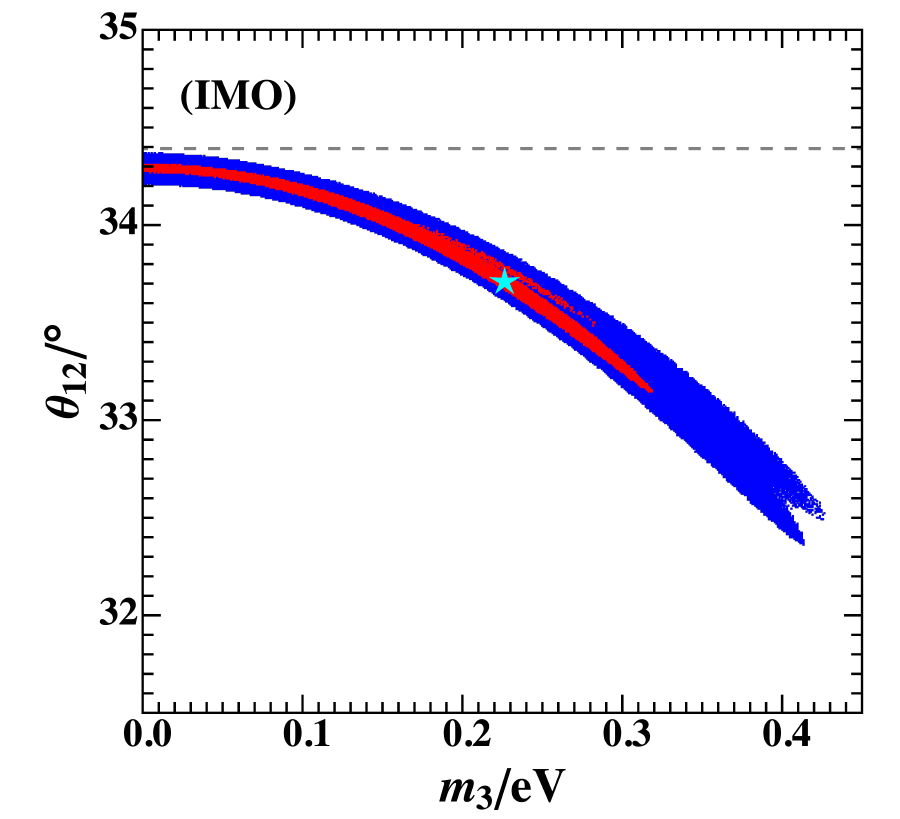} 
		%		\caption{}
		%		\label{fig:sub-d-grid}
	\end{subfigure}
	\hfill
	\begin{subfigure}{0.32\textwidth}
		\centering
		\includegraphics[width=\linewidth]{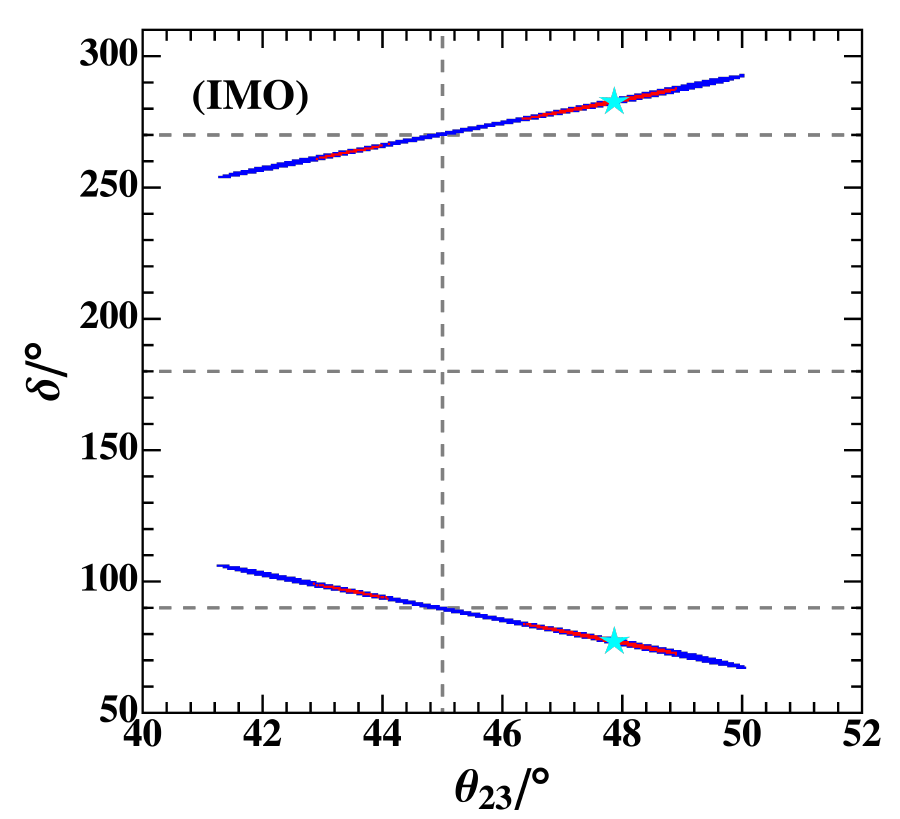}
		%		\caption{}
		%		\label{fig:sub-b-grid}
	\end{subfigure}
	\caption{Correlations between $\left( \theta^{}_{12}, \theta^{}_{13} \right) $, $\left( \theta^{}_{12}, m^{}_{\rm lightest} \right) $, $\left( \delta, \theta^{}_{23} \right) $ pairs at $\Lambda^{}_{\rm EW}$ in the Dirac case. The upper row is for the NMO case, and the lower one is for the IMO case. The cyan star, and red and blue regions correspond to $\chi^2 = \chi^2_{\rm min}$, $\Delta \chi^2 \leq 2.30$, and $\Delta \chi^2 \leq 11.83$, respectively.}
	\label{fig:dirac}
\end{figure}

We perform the same numerical analysis in the Dirac case, where the Dirac CP-violating phase $\delta$ is the only physical phase, and all other phases are unphysical and do not affect physical observables. The initial inputs at $\Lambda$ that minimize $\chi^2$, together with the corresponding predictions for the observables at $\Lambda^{}_{\rm EW}$ and minimum value $\chi^2_{\rm min}$ are summarized in Table~\ref{tab:best-fits}. The allowed regions for the parameter pairs $\left( \theta^{}_{12}, \theta^{}_{13} \right) $, $\left( \theta^{}_{12}, m^{}_{\rm lightest} \right) $, $\left( \delta, \theta^{}_{23} \right) $ are shown in Fig.~\ref{fig:dirac}. It shows that the TM1 mixing pattern can indeed coincide with the experimental data very well with $\chi^2_{\rm min} \sim \mathcal{O} \left( 10^{-2} \right)$, and the results for the $\left( \theta^{}_{12}, \theta^{}_{13} \right) $, $\left( \theta^{}_{12}, m^{}_{\rm lightest} \right) $, $\left( \delta, \theta^{}_{23} \right) $ correlations are qualitatively similar to those in the Majorana case without $\rho^\prime$ and $\sigma^\prime$. Thus, most discussions in the Majorana case remain applicable here and will not be repeated. 

It is worth noting that the lightest neutrino mass in the Dirac case spans a wider range than that in the Majorana case. As indicated in Eq.~\eqref{eq:fun-app}, to reach the same value of $f \left( \Lambda^{}_{\rm EW} \right)$ in the Majorana and Dirac cases, one requires $\zeta^{-1}_{21} = \xi^{}_{21}$, implying a larger $m^{}_{\rm lightest}$ in the Dirac case unless $m^{}_1 = 0$ in both cases. Although $m^{}_{\rm lightest}$ in the Dirac case can be larger, the effective neutrino mass $m^{}_\beta$ almost remains below the most stringent constraint from the KATRIN experiment. Moreover, the constraint from neutrinoless double beta decay experiments is absent in the Dirac case. Therefore, the current experimental data exhibit a stronger preference for the TM1 mixing pattern in the Dirac case than in the Majorana case.

\section{Brief Discussion on TM2 Mixing Pattern}\label{sec:TM2}

\begin{figure}[t!]
	\centering
	\begin{subfigure}{0.32\textwidth}
		\centering
		\includegraphics[width=\linewidth]{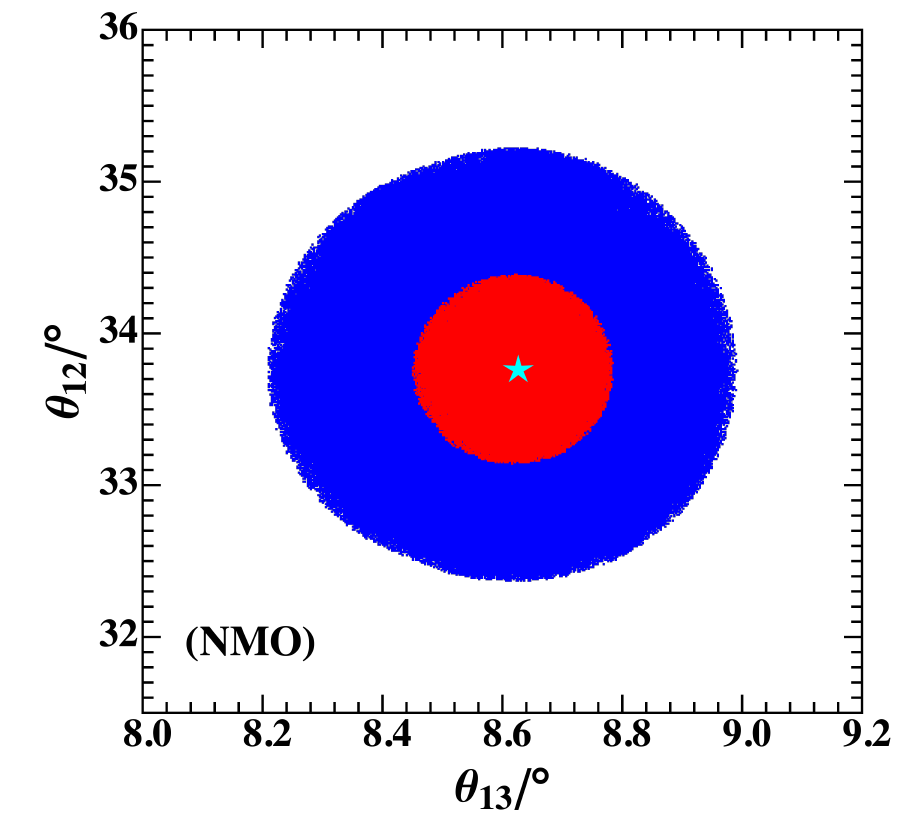}
		%		\caption{}
		%		\label{fig:sub-a-grid}
	\end{subfigure}
	\hfill
	\begin{subfigure}{0.32\textwidth}
		\centering
		\includegraphics[width=\linewidth]{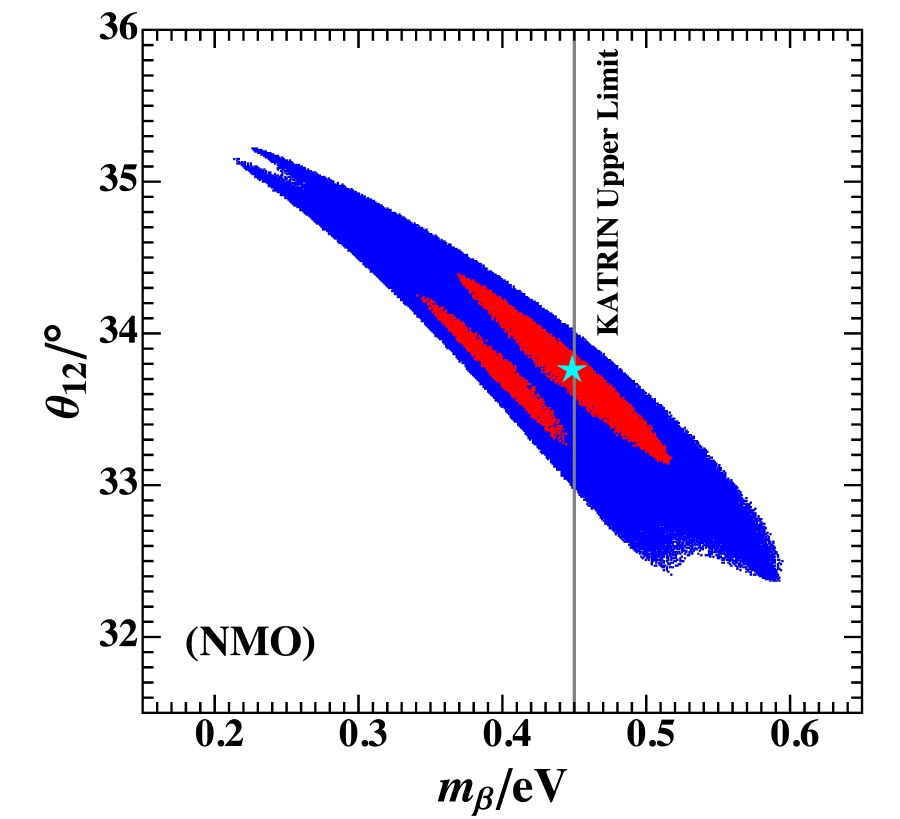} 
		%		\caption{}
		%		\label{fig:sub-d-grid}
	\end{subfigure}
	\hfill
	\begin{subfigure}{0.32\textwidth}
		\centering
		\includegraphics[width=\linewidth]{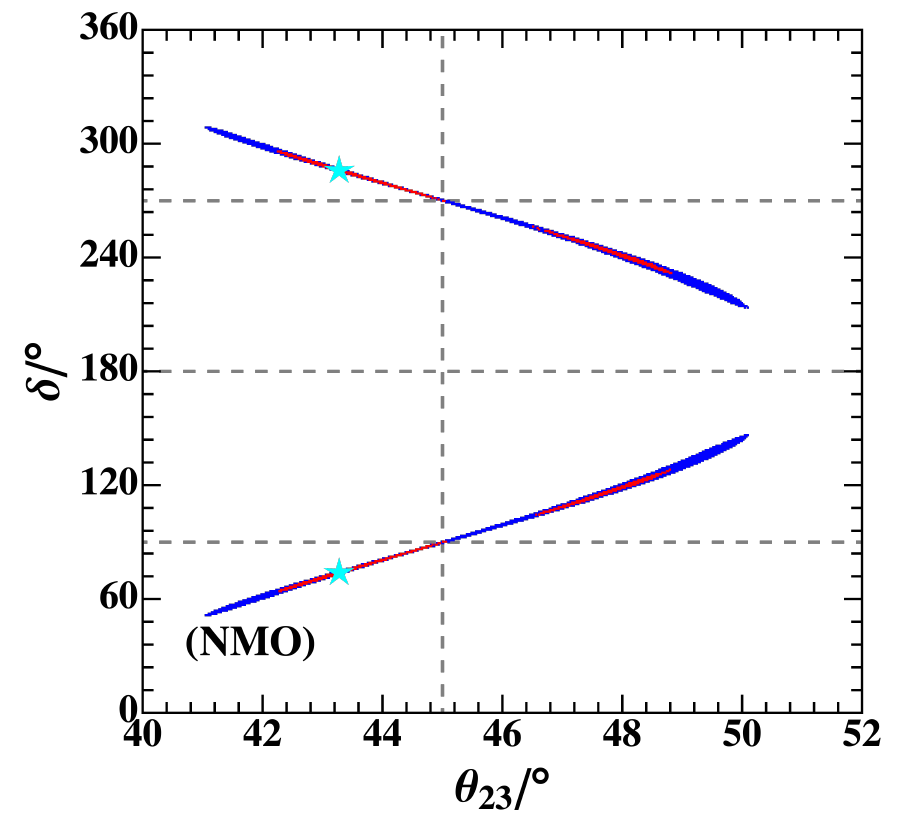}
		%		\caption{}
		%		\label{fig:sub-b-grid}
	\end{subfigure}
	\\
	\begin{subfigure}{0.32\textwidth}
		\centering
		\includegraphics[width=\linewidth]{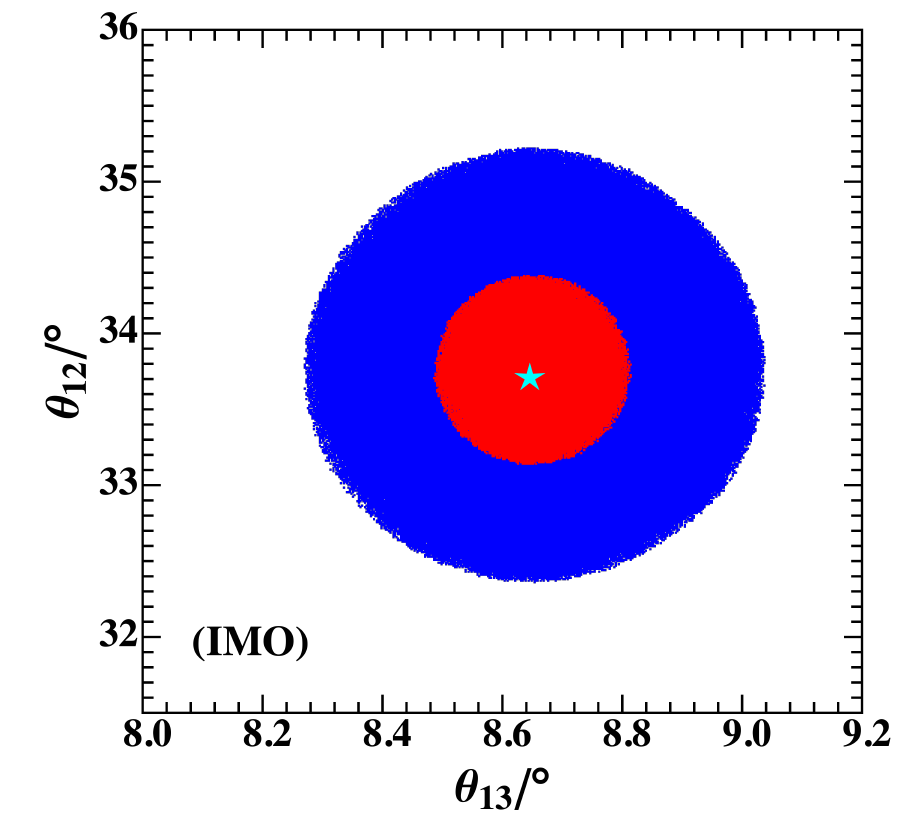}
		%		\caption{}
		%		\label{fig:sub-a-grid}
	\end{subfigure}
	\hfill
	\begin{subfigure}{0.32\textwidth}
		\centering
		\includegraphics[width=\linewidth]{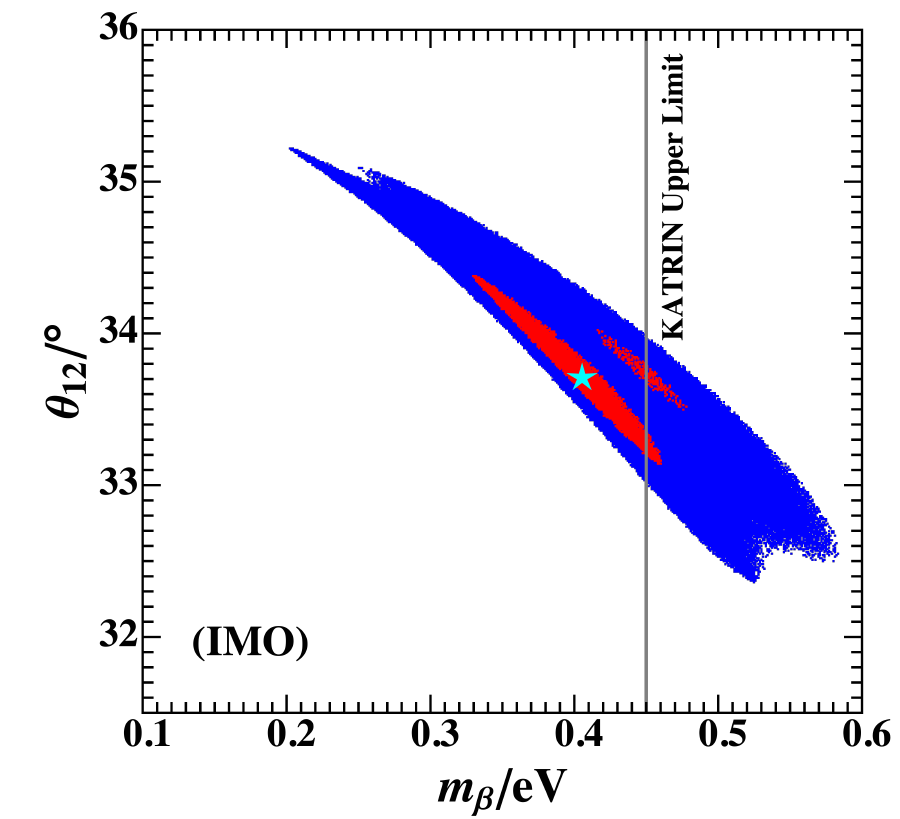} 
		%		\caption{}
		%		\label{fig:sub-d-grid}
	\end{subfigure}
	\hfill
	\begin{subfigure}{0.32\textwidth}
		\centering
		\includegraphics[width=\linewidth]{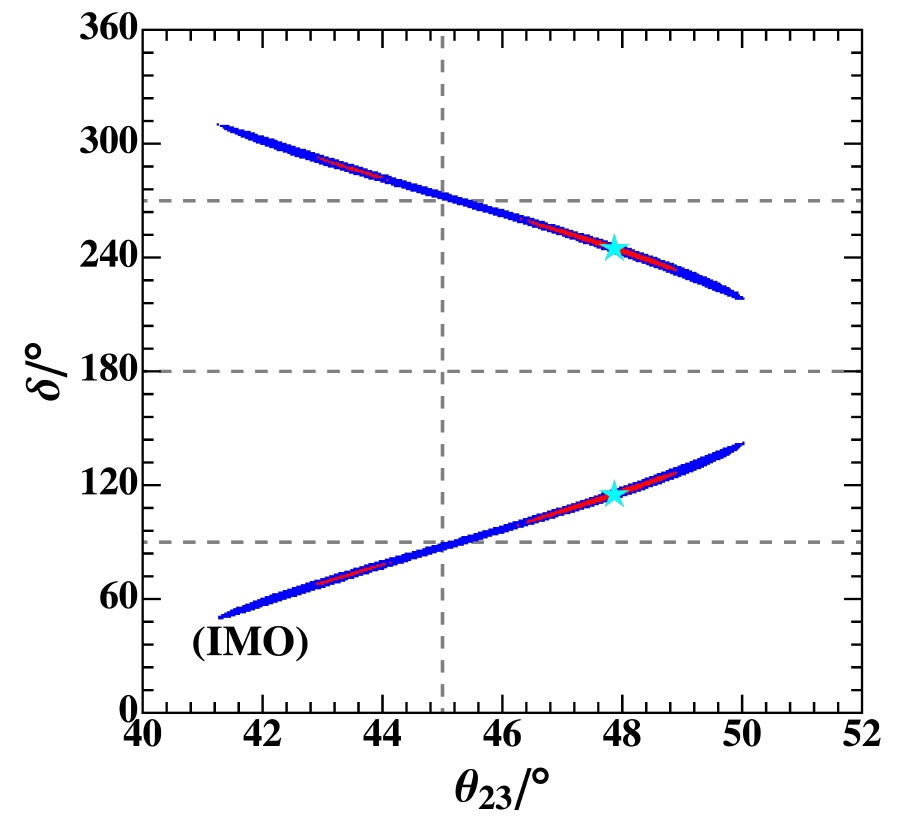}
		%		\caption{}
		%		\label{fig:sub-b-grid}
	\end{subfigure}
	\caption{Correlations between $\left( \theta^{}_{12}, \theta^{}_{13} \right) $, $\left( \theta^{}_{12}, m^{}_\beta \right) $, $\left( \delta, \theta^{}_{23} \right) $ parameter pairs at $\Lambda^{}_{\rm EW}$ in the Dirac case with the TM2 mixing pattern. The upper row is for the NMO case, and the lower one is for the IMO case. The cyan star, and red and blue regions correspond to $\chi^2 = \chi^2_{\rm min}$, $\Delta \chi^2 \leq 2.30$, and $\Delta \chi^2 \leq 11.83$, respectively.}
	\label{fig:dirac-TM2}
\end{figure}

As explicitly illustrated in Fig.~\ref{fig:correlation}, the TM2 mixing pattern without RG corrections already lies outside the experimental $3\sigma$ region of $\theta^{}_{12}$ and $\theta^{}_{13}$ once the latest JUNO result is taken into account. Compared with the TM1 mixing pattern, larger RG corrections (or equivalently, a larger $m^{}_{\rm lightest}$) are required to make the TM2 pattern consistent with experimental data. Consequently, the TM2 mixing pattern in the Majorana case faces even stronger constraints from neutrinoless double beta decay experiments. Indeed, numerical calculations show that the entire allowed range of $m^{}_{\beta\beta}$ for both neutrino mass hierarchies in the Majorana case lies above the KamLAND-Zen upper limit, even if two additional phases $\rho^\prime$ and $\sigma^\prime$ are introduced. This means that the TM2 mixing pattern in the Majorana case is essentially excluded by current experimental data, even after including RG corrections. In contrast, substantially viable parameter regions still exist in the Dirac case, thanks to the absence of constraints from neutrinoless double beta decay experiments. 

The allowed regions of $\left( \theta^{}_{12}, \theta^{}_{13} \right)$, $\left( \theta^{}_{12}, m^{}_\beta \right)$, and $\left( \delta, \theta^{}_{23} \right)$ parameter pairs in the Dirac case are presented in Fig.~\ref{fig:dirac-TM2}, where the minimum value $\chi^2_{\rm min}$ is $9.4\times 10^{-3}$ and $6.6 \times 10^{-3}$ in the NMO and IMO cases, respectively. Some further discussions are given as below:
\begin{itemize}
	\item Unlike the results for the TM1 mixing pattern, the allowed regions of $\theta^{}_{12}$ and $\theta^{}_{13}$ in the two left panels of Fig.~\ref{fig:dirac-TM2} cover their entire $1\sigma$ and $3\sigma$ regions allowed by current experimental data. This behavior originates from the fact that the TM2 $\theta^{}_{12}$-$\theta^{}_{13}$ relation without RG corrections lies outside the experimental $3\sigma$ region, as shown in Fig.~\ref{fig:correlation}. Moreover, the gap between the TM2 $\theta^{}_{12}$-$\theta^{}_{13}$ relation and the experimentally allowed region gives rise to a lower limit on the lightest neutrino mass $m^{}_{\rm lightest}$. Therefore, $m^{}_{\rm lightest}$ is constrained with both upper and lower bounds, namely, $ 0.21~{\rm eV}~ \lesssim m^{}_1 \lesssim 0.59~{\rm eV}$ in the NMO case and $ 0.20~{\rm eV}~ \lesssim m^{}_3 \lesssim 0.58~{\rm eV}$ in the IMO case. These allowed ranges will be further narrowed as the precision of the JUNO measurement on $\theta^{}_{12}$ continues to improve.
	
	\item The two right panels of Fig.~\ref{fig:dirac-TM2} display the allowed regions of $\delta$ and $\theta^{}_{23}$. It can be seen that the quadrant of $\delta$ is inversely correlated with the octant of $\theta^{}_{23}$, in contrast to the correlation found in the TM1 mixing pattern. More specifically, the first and fourth quadrants of $\delta$ correspond to the first octant of $\theta^{}_{23}$, while the second and third quadrants of $\delta$ correspond to the second octant of $\theta^{}_{23}$. This difference between the TM1 and TM2 mixing patterns arises from their distinct $\delta$-$\theta^{}_{23}$ relations. The TM1 $\delta$-$\theta^{}_{23}$ relation is given in Eq.~\eqref{eq:TM1-relations}, and the one in the TM2 pattern is found to be
	\begin{eqnarray}
		\cos\delta = \frac{\cot2\theta^{}_{23} \left( 1 - 2\sin^2\theta^{}_{13} \right) }{\sin\theta^{}_{13} \sqrt{2-3\sin^2\theta^{}_{13}}} \;.
	\end{eqnarray}
	Obviously, $\cos\delta$ takes opposite signs in the TM1 and TM2 mixing patterns, given that the value of $\theta^{}_{13}$ is small.
	
	\item The two middle panels of Fig.~\ref{fig:dirac-TM2} show the allowed region of $\theta^{}_{12}$ and $m^{}_\beta$, together with the KATRIN upper limit. In the NMO case, the best-fit point and nearly half of $1\sigma$ and $3\sigma$ regions are below the KATRIN bound, and in the IMO case, the best-fit point and most $1\sigma$ region remain below the limit. Therefore, the current experimental data show a slight preference for the IMO case over the NMO case. In addition, if the KATRIN experiment achieves its final sensitivity of sub-300~meV~\cite{KATRIN:2024cdt} without any discovery, most of the currently allowed parameter space including the entire $1\sigma$ regions will be excluded in both the NMO and IMO cases.
\end{itemize}

Before ending this section, we briefly comment on cosmological constraints on the sum of neutrino masses. Within the $\Lambda$CDM + $\sum m^{}_i$ framework, the combination of cosmic microwave background and baryon acoustic oscillation data has yielded an increasingly stringent upper limit of $\sum m^{}_i \lesssim 0.07$ eV at the 95\% confidence level~\cite{Planck:2018vyg,DESI:2024mwx,DESI:2025zgx}. Notably, this bound lies below the minimal value $\sum m^{}_i \simeq 0.1$ eV allowed by neutrino oscillation data in the IMO case and is close to the minimal value $\sum m^{}_i \simeq 0.06$ eV in the NMO case. Furthermore, a negative total neutrino mass appears statistically favored when the physical prior $\sum m^{}_i >0$ is relaxed (see, e.g., Refs.~\cite{Craig:2024tky,Green:2024xbb,Elbers:2024sha,Naredo-Tuero:2024sgf,Jiang:2024viw}). Indeed, these cosmological bounds depend sensitively on datasets included in the analysis, analysis methodology, and cosmological assumptions, and can be relaxed to the sub-eV scale in extended scenarios, such as models with dynamical dark energy ~\cite{ParticleDataGroup:2024cfk,Capozzi:2025wyn,Planck:2018vyg,DESI:2024mwx,DESI:2025zgx,Elbers:2025vlz,Chebat:2025kes,DESI:2025ffm}. From this perspective, the TM2 mixing pattern that requires $0.2~{\rm eV} \lesssim m^{}_{\rm lightest} \lesssim 0.6~{\rm eV}$ is robustly disfavored by current cosmological observations, whereas the TM1 mixing pattern remains consistent with oscillation data at the $1\sigma$ level. Nevertheless, the relatively large mass scale $m^{}_{\rm lightest} \sim 0.15$ eV (or $0.23$ eV) required by the best-fit point of the TM1 mixing pattern in the Majorana (or Dirac) case is in significant tension with the most restrictive cosmological limits within the minimal $\Lambda$CDM or its simple extensions. This tension appears difficult to resolve through standard cosmological modifications alone. However, if one takes into account threshold effects potentially enhancing RG corrections to the mixing angles in models with non-degenerate heavy particles~\cite{Antusch:2005gp,Antusch:2002rr,Zhang:2024weq}, it is possible to reduce the required neutrino mass scale and thereby alleviate its tension with the cosmological constraints.

\section{Summary}\label{sec:Summary}

The JUNO experiment has very recently released its first measurement results based on 59.1 days of data, and provided a world-leading determination of $\theta^{}_{12}$ and $\Delta m^2_{21}$. This improvement in the precision of $\theta_{12}$ poses a greater challenge to various neutrino mass models or flavor mixing patterns that predict a specific value of $\theta_{12}$ or establish correlations of $\theta^{}_{12}$ with other mixing parameters. Against this background, we have investigated the impact of the latest JUNO results on the TM1 and TM2 mixing patterns, which are two popular variants of the tri-bimaximal mixing pattern and give intriguing correlations between $\theta^{}_{12}$ and $\theta^{}_{13}$. These TM1 and TM2 $\theta^{}_{12}$-$\theta^{}_{13}$ correlations were consistent with the NuFIT 6.0 global analysis of all available experimental data prior to JUNO's results within the $1\sigma$ and $3\sigma$ levels, respectively, but now the former sits on the edge of the allowed $1\sigma$ region while the latter lies outside of the allowed $3\sigma$ region once the latest JUNO measurement is included. 

To explore the possibility of reconciling the TM1 and TM2 mixing patterns with experimental data, we have further investigated their RG running effects from a typical super-high energy scale down to the electroweak scale, assuming no additional physics thresholds in between, in both Majorana and Dirac neutrino cases. Through detailed analytical and numerical analyses, we have demonstrated that RG corrections can significantly modify the $\theta^{}_{12}$-$\theta^{}_{13}$ correlations and, fortunately, bring them into perfect agreement with the latest JUNO measurement when neutrino masses are nearly degenerate. However, in the Majorana case, this required nearly degenerate neutrino mass spectrum is strongly constrained by the upper limit on the effective neutrino mass $m^{}_{\beta\beta}$ probed in neutrinoless double beta decay experiments, while in the Dirac case such constraints are absent. Consequently, the TM mixing patterns with Dirac neutrinos receive more support from both oscillation and non-oscillation experiments than those with Majorana neutrinos. The TM1 mixing pattern in the Majorana case still retains viable parameter space that lies below the KamLAND-Zen upper limit or within its uncertainty for both neutrino mass orderings, particularly after introducing two additional rephasing phases. By contrast, almost the entire parameter space of the TM2 mixing pattern in the Majorana case has been excluded by the KamLAND-Zen limit for both neutrino mass orderings. Furthermore, among the two TM mixing patterns, current experimental data clearly favor TM1 over TM2, regardless of whether RG corrections are included. This point in the Majorana case has been clearly described above. In the Dirac case, although the constraints from neutrinoless double beta decay experiments are absent, the parameter space of the TM2 mixing pattern is partially excluded by the current KATRIN upper limit and could be essentially ruled out once the KATRIN experiment reaches its final sensitivity without any discovery. In contrast, the TM1 mixing pattern remains safely below both the present and projected KATRIN limits.

Finally, we note that our results ignoring small parameter correlations in the defined $\chi^2$ function in Eq.~\eqref{eq:chi-fun} serve as a robust first approximation, as a fully comprehensive analysis incorporating all exact multidimensional parameter correlations is beyond the scope of the present work. Nevertheless, with more forthcoming data expected from JUNO and other oscillation and non-oscillation experiments, the precision of lepton flavor mixing parameters will be further improved, and additional information on the absolute neutrino masses will be obtained. These advances will not only provide decisive tests of the two attractive mixing patterns but also offer valuable insights into the underlying lepton flavor structure.

\section*{Acknowledgements}

This work was partially supported by the Collaborative Research Center SFB1258 and the Deutsche Forschungsgemeinschaft (DFG, German Research Foundation) under Germany's Excellence Strategy - EXC-2094 - 390783311.

\end{document}